\documentclass[a4paper]{article}

\usepackage{vmargin}   
\setmarginsrb{3cm}{2cm}{3cm}{2cm}{1cm}{1cm}{1cm}{1cm}   
\usepackage{amsmath}   
\usepackage{amssymb}  
\usepackage[dvips]{graphicx}  
\usepackage{rotating}   
\usepackage{psfrag} 
\usepackage{fancyhdr}
\usepackage{pstricks}
\newcommand{\be}{\begin{equation}}
\newcommand{\ee}{\end{equation}}
\newcommand{\ba}{\begin{eqnarray}}
\newcommand{\ea}{\end{eqnarray}}
\newcommand{\st}{\scriptstyle}

\newcommand{\la}{\langle}
\newcommand{\ra}{\rangle}

\begin{document}

\title{Spin-Foam Models and the Physical Scalar Product}
\author{\large
Emanuele Alesci${}^{1ab}$, Karim Noui${}^{2c}$ , Francesco Sardelli${}^{3cd}$
 \\[3mm]
\em
\normalsize
{${}^a$Laboratoire de Physique, ENS Lyon, CNRS UMR 5672, 46 All\'ee d'Italie, 69007 Lyon, France EU}
\\ \em\normalsize{${}^b$Centre de Physique Th\'eorique de Luminy%
\footnote{Unit\'e mixte de recherche (UMR 6207) du CNRS et des
Universit\'es de Provence (Aix-Marseille I), de la Mediterran\'ee
(Aix-Marseille II) et du Sud (Toulon-Var); laboratoire affili\'e \`a
la FRUMAM (FR 2291).}, Universit\'e de la M\'editerran\'ee, F-13288 Marseille EU}
\\ \em\normalsize{${}^c$Laboratoire de Math\'ematiques et de Physique Th\'eorique\footnote{
UMR/CNRS 6083, F\'ed\'eration Denis Poisson\newline
\hspace{2mm} e-mail: \,\, ${}^1$alesci@fis.uniroma3.it \, ,\,\,  ${}^2$noui@lmpt.univ-tours.fr \, , 
\,\,${}^3$francesco.sardelli@yahoo.it},
Parc de Grammont, 37200 Tours EU}
\\\em\normalsize{${}^d$} Dipartimento di Fisica Enrico Fermi, Universit\`a di Pisa, 56127 Pisa EU}

\date{\small\today}

\maketitle \vspace{-.6cm}  

\begin{abstract}
\noindent This paper aims at clarifying the link between Loop Quantum Gravity
and Spin-Foam models in four dimensions. Starting from the canonical framework,
we construct an operator $P$ acting
on the space of cylindrical functions $\text{Cyl}(\Gamma)$, where $\Gamma$ is 
the 4-simplex graph, such that its matrix elements are, up to some normalization factors,
the vertex amplitude of Spin-Foam models. The Spin-Foam models we are considering are the 
topological model, the Barrett-Crane model and the Engle-Pereira-Rovelli model.  
The operator $P$ is usually called the ``projector" into physical states and its matrix elements
gives the physical scalar product. Therefore, we relate the physical scalar product of Loop Quantum
Gravity to vertex amplitudes of some Spin-Foam models.
We discuss the possibility to extend the action of $P$ to any cylindrical functions
on the space manifold.
\end{abstract}

\section*{Introduction}
Finding the physical scalar product is certainly one of the most important
question of Loop Quantum Gravity \cite{lqg,lqg2}. This is somehow equivalent to the problem
of finding solutions of the remaining scalar constraint which is, so far, still
an open issue. Two main and very active directions have been followed to tackle the problem:
(i) formulating consistently the scalar constraint as a well-defined operator
acting on the kinematical Hilbert space; (ii) making sense of the covariant
quantization to compute physical transitions amplitudes between states of quantum
geometry.  The former has been explored mainly by Thiemann\cite{thiemann} and collaborators:
very tricky and very nice regularizations of the scalar constraints have been found; 
the important question is now to extract physical solutions out of it.
The master constraint program \cite{master} has been considered to that aim. 
Spin-Foam models \cite{spinfoam} are the covariant alternative attempt to solve the problem:
they propose a way to ``compute" the path integral of gravity where space-time
appears as a combinatorial foam which can be understood as a covariant 
generalization of the notion of spin-networks. Then a spin-foam is 
somehow interpreted as the structure which encodes the ``time evolution"
of a state of quantum gravity. Spin-Foam models have been studied intensively these
last years to answer some fundamental questions they have raised, 
two of the most important being the following: 
What is the precise link between Spin-Foam models and the path integral of quantum gravity?
Can we establish an explicit link between Spin-Foam models and Loop Quantum Gravity as in the three dimensional case \cite{NP} ?

To understand the meanning of the first question,
it is worth recalling that Spin-Foam models are only ansatz for the path integral of quantum gravity. 
The ansatz is based
on the Plebanski formulation of general relativity \cite{plebanski} where gravity appears as a topological BF theory
supplemented with simplicity constraints on the B field. The path integral of 
a (Euclidean) BF theory is a topological invariant which can be reformulated ``exactly"
as a Spin-Foam model which is called, in a more mathematical language, a state sum model.
The natural idea is to try to impose the simplicity constraints at the level of the path integral
to get a Spin-Foam model for gravity. Barrett and Crane (BC) \cite{BC} proposed a first model: it was studied a lot
but recently it was shown not to reproduce expected behavior at the semi-classical limit \cite{difficulties} while computing
the 2-points functions of gravity in the context of LQG propagator calculations \cite{propagator}. 
It was then realized that the way Barrett and Crane had imposed the simplicity constraints at
the level of the Spin-Foam would have been, in a sense, too strong.
Engle, Pereira and Rovelli (EPR) have proposed a new model \cite{EPR} which seems a more promising candidate:
in a subsequent paper with Livine \cite{ELPR}, they have proposed a way to impose the simplicity constraints using
 the ``master constraint" techniques 
introduced in the context of canonical quantization by Thiemann.
One can incorporate the Immirzi parameter in the new model and it is possible to extend it
 to the Lorenzian case\cite{EPRlorentz}. In the meanwhile another model from Freidel and Krasnov (FK) \cite{FK} has appeared. The FK model instead imposes the constraints using the coherent states techniques introduced by 
Livine and Speziale \cite{LS}. All these new models are under study at this moment \cite{CF} in order to see, 
in particular, if they behave correctly in the classical and semi-classical limits \cite{EC}. 

The second question concerning the link between canonical and covariant quantizations
of gravity has been quite problematic for a long time: the Lorentzian BC model seemed 
incompatible with Loop Quantum Gravity because it is known that the spectra of the area operator are not identical
in the two approaches. 
Covariant Loop Quantum Gravity \cite{Alexandrov} was introduced to repair this problem
modifying (in a covariant way) the canonical quantization: the obtained theory is unfortunately
too cumbersome to be useful for the moment. Instead of modifying the canonical quantization, one
could consider standard Loop Quantm Gravity as the good framework for the canonical quantization
of gravity and think about finding a Spin-Foam model consistent with this approach.
This is exactly what the new EPR model is doing: the projected states of the new model are the
standard spin-network states; the spectrum of the area operators in the covariant quantization is the
same as the one in the canonical quantization. Therefore, the EPR model seems to be a good candidate
to test if Spin-Foam models can explicitely realize a ``projection" (in the sense of Loop Quantum Gravity)
into physical states. Indeed, we expect the physical scalar product between two spin-network
states to be given by the Spin-Foam amplitude associated to a graph whose boundaries are the
two given spin-networks.

\medskip

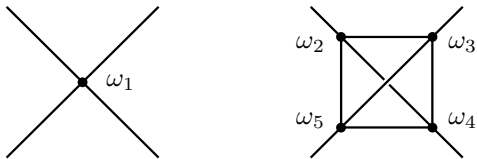
\begin{figure}[h]
\begin{center}
\ifx\JPicScale\undefined\def\JPicScale{1}\fi
\psset{unit=\JPicScale mm}
\psset{linewidth=0.3,dotsep=1,hatchwidth=0.3,hatchsep=1.5,shadowsize=1,dimen=middle}
\psset{dotsize=0.7 2.5,dotscale=1 1,fillcolor=black}
\psset{arrowsize=1 2,arrowlength=1,arrowinset=0.25,tbarsize=0.7 5,bracketlength=0.15,rbracketlength=0.15}
\begin{pspicture}(0,0)(63,20)
\psline(3,20)(23,0)
\psline(23,20)(3,0)
\rput{0}(13,10){\psellipse[fillstyle=solid](0,0)(0.5,-0.5)}
\rput(18,10){$\omega_1$}
\psline(43,20)(63,0)
\psline[border=0.3](63,20)(43,0)
\rput{0}(47,4){\psellipse[fillstyle=solid](0,0)(0.5,-0.5)}
\rput(43,15){$\omega_2$}
\rput{0}(59,4){\psellipse[fillstyle=solid](0,0)(0.5,-0.5)}
\rput{0}(47,16){\psellipse[fillstyle=solid](0,0)(0.5,-0.5)}
\rput{0}(59,16){\psellipse[fillstyle=solid](0,0)(0.5,-0.5)}
\psline(47,16)(47,4)
\psline(47,4)(59,4)
\psline(59,4)(59,16)
\psline(59,16)(47,16)
\rput(63,15){$\omega_3$}
\rput(63,5){$\omega_4$}
\rput(43,5){$\omega_5$}
\end{pspicture}

\caption{\small{Illustration of the 1-tetrahedron state $\tau_1$ on the left and the 4-tetrahedron state 
$\tau_4$ on the right. Vertices, labelled by $i\in\{0,5\}$, are colored with intertwiners $\omega_i$
and edges $\ell_{ij}$ with representations $I_{ij}$. The 4 free ends are colored with representations
$I_{1i}$.}}
\label{tau}
\end{center}
\end{figure}

This article aims at clarifying this relation with a simple example.
We consider Euclidean  Spin-Foam models associated to the group
$G=SU(2)\times SU(2)$. It is characterized by its vertex amplitude $V$: 
the vertex amplitude is the weight associated to a 4-simplex;
it is therefore a function $V(I_{ij},\omega_i)$ of the $G$-representations $I_{ij}$
coloring the 10 faces of the 4-simplex and of the $G$-intertwiners $\omega_i$ associated
to the 5 tetrahedra of the 4-simplex. The index $i$ runs from 1 to 5 and labels the five
tetrahedra in the boundary of the 4-simplex. 
We want to interpret this vertex amplitude $V$ as the physical scalar product between
two spin-network states: the 1-tetrahedron state $\tau_1$ and the 4-tetrahedra state $\tau_4$
associated to spin-networks respectively dual to one tetrahedron and to four tetrahedra
as illustrated in the figure (\ref{tau}). The free ends of these spin-networks coincinde and therefore $\tau_1$
and $\tau_4$ are particular cylindrical functions of the same graph, denoted
$\tilde{\Gamma}$, as illustrated in the figure (\ref{Gammaema}) in the core of the paper. The graph $\tilde{\Gamma}$
is the union of the 4-simplex graph $\Gamma$ with four free edges and it was introduced to take into account
the free ends of the states $\tau_1$ and $\tau_4$. 

More precisely, we construct on operator $P$ acting on the space of cylindrical functions $\text{Cyl}(\tilde{\Gamma})$ 
such that its matrix elements are related to the vertex amplitude of Spin-Foam models as follows:
\be\label{general projector}
\la \tau_4 , P \tau_1 \ra \; = \; N \,V(I_{ij},\omega_i) 
\ee
where $N$ is an eventual normalization factor. In that sense, the matrix element $\la \tau_4 , P \tau_1 \ra$
would be the physical scalar product between the kinematical states $\tau_1$ and $\tau_4$.
In fact, the bra-ket notation for the physical scalar product might be misleading because
mathematically $P$ is a linear form on the space $\text{Cyl}(\tilde{\Gamma})$, 
i.e. $P \in \text{Cyl}(\tilde{\Gamma})^*$, abusively called a ``projector",
and the physical scalar product is $\la \tau_4 , P\tau_1 \ra=P(\overline{\tau_4}\tau_1)$. 
In the context of Gelfand-Naimark-Segal theory (see the Ashtekar-Lewandowski review \cite{lqg} and references therein),
$P$, if it satisfies some additional properties, would be a state and would allow to construct the whole 
physical Hilbert space in principle.
  
We find a solution for the projector $P$ for different Spin-Foam models: the topological $SU(2)$ BF model
whose vertex $V_{BF}$ is the 15j symbol of $SU(2)$ (this system has no physical relevence); 
the BC model whose vertex $V_{BC}$ is the well-known 10j symbol; the new model whose vertex
$V_{EPR}$ has been defined recently and also the FK model whose vertex construction is a direct extension
of the EPR one (in this paper we concentrate only on the vertex amplitude without discussing the measure factors associated to the FK model, see \cite{FK}). The projector $P_{BF}$ associated to the topological model
is a multiplicative operator which acts only on the edges of the spin-networks and imposes
that the connection is flat. The projectors $P_{BC}$ and $P_{EPR}$ respectively associated to the BC and the 
EPR models act both on the vertices (as derivative operators, in the sense that it involves left and right invariant
derivatives) and on the edges of spin-networks. 
Note that we construct one solution of $P$ and we do not precisely address the question of the unicity 
in this article.

\medskip

The plan of this article is the following. In Section 1, we propose a simple and
general integral formula of the vertex
amplitudes of  Euclidean 4 dimensional Spin-Foam models. It is quite an universal formula for it contains
as particular cases the vertices of all the known models as the topological, the BC and the EPR models.
In Section 2, we make use of this formula to construct physical operators for each model in a way similar
to the 3 dimensional case. 
More precisely, we find a solution to the equation (\ref{general projector})
 for each model and we discuss the properties of these solutions. 
We conclude with some perspectives.

\section{The vertex of a Spin-Foam model}
In this Section, we present some properties concerning the vertex amplitude of several Spin-Foam
models. The notion of vertex amplitude is defined in the first part where we give a very brief introduction
on Spin-Foam models. In a second part, we propose a general and rather simple integral formula for the 
vertex amplitude which will be useful in the next Section to make a link with the canonical quantization. In the last part
of this Section, we illustrate this formula in the particular models we are interested in, namely the topological,
the BC and the EPR models.
Furthermore, we underline that the BC and the EPR models are particular cases of a large
class of Spin-Foam models. We present the construction of this class of Spin-Foam models and we show that
their vertex amplitude admits an integral formulation of the same type.

\subsection{A brief introduction on Spin-Foam models}
A Spin-Foam model is basicly the assignment of a complex amplitude ${\cal A}(\cal T)$ to any triangulation 
$\cal T$ of a given
four dimensional manifold $\cal M$. The triangulation consists in the union $\cup_{i=2}^4 {\cal T}_i$
of the set of its faces ${\cal T}_2$, the set of its tetrahedra ${\cal T}_3$ and the set of its 4-simplices
${\cal T}_4$.
The amplitude $\cal A$ is constructed from the representation theory of a 
given Lie group $G$ that we assume compact for simplicity. 
To do so, one first colors each face $f \in {\cal T}_2$ 
with an unitary irreducible representation (UIR) $j_f$ of $G$ and each tetrahedron $t \in {\cal T}_3$ 
with intertwiners $\iota_t$ 
between representations coloring its 
four faces. Then, one associates an amplitude ${\cal A}_2(j_f)$ to each face $f$, an amplitude 
${\cal A}_3(\omega_t,j_{f_t})$
to each tetrahedron $t$ which depends on the intertwiner $\omega_t$ and on the representations coloring its 4
faces $f_t$, 
and an amplitude $V(\omega_{t_s},j_{f_s})$ to each 4-simplex $s$ which depends on
the representations $j_{f_s}$ and $\omega_{t_s}$ coloring its 10 faces $f_s$ and 5 tetrahedra $t_s$. 
Finally, the spin-foam
amplitude is formally defined by the series
\begin{eqnarray}\label{generalSFamp}
{\cal A}({\cal T}) \; \equiv \; \sum_{\{j_f\},\{\omega_t\}} \prod_{f \in {\cal T}_2} {\cal A}_2(j_f) \, 
\prod_{t \in {\cal T}_3} {\cal A}_3(\omega_{t},j_{f_t}) \, \prod_{s \in {\cal T}_4} V(\omega_{t_s},j_{f_s})
\end{eqnarray}
where the sum runs into a certain subset of UIR and intertwiners of $G$. 
The sum is a priori infinite and therefore the amplitude is only defined formally at this stage
unless it is convergent. Notice that in all the models that have been studied in the literature,
the amplitude ${\cal A}_3$ is assumed to depends on the intertwiners $\omega_t$ only.
The function $V$ is precisely the vertex amplitude of the Spin-Foam model.
To finish with this brief introduction of Spin-Foam models, let us mention
that the previous construction could be generalized to the case where $G$ is non-compact and to the
case where $G$ is replaced by a quantum group. Spin-Foam models can also be defined for any dimensional
manifold $\cal M$.

\medskip

In this paper, we consider exclusively the case where $\cal M$ is 4-dimensional and we study 
some properties of the vertex amplitude $V$ only. Therefore, we will not mention the amplitudes
${\cal A}_2$ and ${\cal A}_3$ when we discuss the Spin-Foam models in the sequel; as a result, we will omit
any discussion concerning the amplitude $\cal A$ and a fortiori the question of its convergence. We hope to study
these aspects in the future.
Furthermore, we will consider Euclidean Spin-Foam models only which are associated to the compact Lie groups
$G=SU(2)$ (for the topological model) or $G=SU(2)\times SU(2)$ (for the BC and EPR models).
Letters $I,J,\cdots$ label unitary irreducible representations of the group $G$ and the associated 
vector spaces are denoted $U_I, U_J, \cdots$. 
When $G=SU(2)$, $I$ is a half-interger whereas it is a couple of half-integers when $G=SU(2)\times SU(2)$.
Due to the compactness of $G$, each representation $I$ 
is finite dimensional and associates to any $g\in G$ a finite dimensional matrix which will be denoted  
$R^I(g)$ when $G=SU(2)\times SU(2)$ and $D^{I}(g)$ in the other case. To a representation $I$ is associated
a contragredient (or a dual) representation $I^*$ such that $R^{I^*}(g)={}^tR^I(g^{-1})$ and the same for the $SU(2)$
representations $D^{I^*}$; it is common to identify $U_I^*\equiv U_{I^*}$ to $U_I$.
More precision concerning the representation theory of the groups
$G$  will be given later.

The vertex $V(I_{ij},\omega_i)$ is then a function of the 5 intertwiners $\omega_i$ coloring the 5 
tetrahedra (which are ordered and labelled by $i \in \{1,5\}$) of a 4-simplex and of the 
10 representations $(I_{ij})_{i<j}$ of $G$ coloring the 10 faces at the intersections of the tetrahedra $i$ and $j$;
$\omega_i: \otimes_{j>i}U_{I_{ij}} \rightarrow \otimes_{j<i}U_{I_{ji}}$ is an intertwiner between the 
representations $I_{ij}$ ``meeting" at the tetrahedron $i$.
In the next part, we are going to show that the vertex amplitude of all the models we consider 
can  be written as an integral over 10 copies of the 3-sphere $S^3$ as follows:
\be\label{general vertex}
V(I_{ij},\omega_i) \; = \; \int \big(\prod_{i<j} dx_{ij} \big)\, C(x_{ij})\,  {\cal V}(I_{ij},\omega_i;x_{ij})  
\ee
where $C(x_{ij})$ is a universal function, in the sense that it is model independent, which reads
\begin{eqnarray}
C(x_{ij}) \; \equiv \; \int \big(\prod_{i=1}^5 dx_i\big) \, \delta(x_{ij}^{-1}x_ix_j^{-1})\;.
\end{eqnarray}
${\cal V}$ is a model dependent function of the variables $x_{ij}$.
As we will see in the next Section, such a formula will be crucial to link Spin-Foam models
with Loop Quantum Gravity.

\subsection{A General expression of the vertex}
There exists many equivalent ways to define the vertex amplitude of a Spin-Foam models.
For our purposes, it is convenient to view 
the vertex amplitude as a ``Feynman graph" evaluation of a closed oriented graph which is dual
to a 4-simplex. The dual of a 4-simplex $\Gamma$ is in fact topologically equivalent to a 4-simplex and then
consists in a set of 5 vertices linked  by 10 edges:
we endow the set of vertices with a linear ordering such that the vertices are labelled with
an integer $i \in \{1,5\}$; this ordering induces a natural orientation on the links,
indeed the link $\ell_{ij}$ between the edges $i$ and $j$ is oriented from $i$ to $j$ if $i<j$.
One associates a complex amplitude to this graph using the following ``Feynman" rules: 
each oriented link $\ell_{ij}$, with $i<j$, is associated to a UIR of $G$ denoted $I_{ij}$ 
(the opposite link $\ell_{ji}$ is associated to the contragredient representation denoted for simplicity 
$I_{ji}=I_{ij}^*$);
each vertex $i$ is associated to an intertwiner $\omega_i:\otimes_{j>i}U_{I_{ij}}\rightarrow \otimes_{j<i}U_{I_{ji}}$. 
As a result, the ``Feynman evaluation" of such a graph is the scalar obtained by contracting the 10 propagators
with the 5 intertwiners and gives the vertex amplitude which formally reads:
\ba\label{general formula}
V(I_{ij},\omega_i) \; = \; \la \bigotimes_{i=1}^5 \omega_i \ra \,
\equiv \, \sum_{\{e_{ij}\}} \prod_{i=1}^5 
\langle \bigotimes_{j<i} e_{ji} \vert \omega_i \vert \bigotimes_{j>i} e_{ij} \rangle 
\ea
where $e_{ij}$ runs over the finite set of a given orthonormal basis of $U_{I_{ij}}$
and we have used the standard bra-ket notation to denote the vectors $\vert e_{ij} \rangle$ of $U_{I_{ij}}$
and the dual vectors $\langle e_{ij} \vert$. In the language of Loop Quantum Gravity, we would say that 
$V(I_{ij},\omega_i)$ is simply the evaluation of the spin-network associated to the colored graph 
$(\Gamma,\{I_{ij},\omega_i\})$ when the connection is flat.

In order to have a more useful formula,
it will be convenient to trivially identify $\omega_i$ with an element of 
$\text{Hom}(\otimes_{j\neq i} U_{I_{ij}},\mathbb C)$
and then to notice that $\omega_i$ is completely caracterized by a vector $v_i \in \otimes_{j\neq i} U_{I_{ij}}^*$. 
These vectors can be written in the form $v_i=\sum_{(a_{ij})} \alpha_i^{(a_{ij})} \otimes_{j\neq i} v_{a_{ij}}$ 
where $(a_{ij})_{j\neq i}$
is a set whose elements label vectors $v_{a_{ij}} \in U_{I_{ij}}$, $\alpha_i^{(a_{ij})}$ are complex numbers
and the sum is finite.
The explicit relation between $\omega_i$ and $v_i$ is the following:
\ba\label{omega}
\omega_i \; = \;  \langle v_i \vert \int dg \, \bigotimes_{j \neq i} R^{I_{ij}}(g) \; \in \;  
\text{Hom}(\bigotimes_{j\neq i} U_{I_{ij}},\mathbb C)
\ea
where we have used the $SU(2)\times SU(2)$ notations for the representations and
$\int dg$ is the Haar measure of $G$. As a result, the vertex amplitude can be reformulated as a multi-integral
over $G$ according to the formula:
\ba
V(I_{ij},\omega_i) \; = \; \sum_{(a_{ij})} \prod_{i=1}^5 \alpha_i^{(a_{ij})} \,
\int (\prod_{i=1}^5 dg_i) \, \langle \otimes_{i<j} v_{a_{ij}}\vert 
 \bigotimes_{i<j} R^{I_{ij}}(g_ig_j^{-1})  \vert \otimes_{i>j} v_{a_{ij}} \rangle
\ea
which can be written in the following more compact well-known form
\be\label{general formula integral}
V(I_{ij},\omega_i) \; = \; \int (\prod_{i=1}^5 dg_i) \,  ( \otimes_{i=1}^5 v_i )\, \cdot
 (\bigotimes_{i<j} R^{I_{ij}}(g_ig_j^{-1}))
\ee
where the dot $\cdot$ denotes the appropriate contraction between the vectors $v_i$ and the matrices of the 
representations.
This vertex amplitude is in fact rather general and caracterizes partially a large class of Spin-Foam models.
It is general because we have for the moment a total freedom in the choice of the representations and the
intertwiners;
it is nonetheless only partial because we do not consider the amplitudes associated to faces and tetrahedra. 
To go further in the study of this amplitude,
we need to recall some basic results on the representation theory of $SU(2)\times SU(2)$.

\subsubsection{Representation theory of $G$: basic results}
Let us start with the group $SU(2)$: its representations are labelled by a half-integer, the spin $I$;
they are finite dimensional of dimension $d_I=2I+1$ and we denote by $\vert I,i\rangle$ with $i \in [-I,I]$ 
the vector of an orthonormal basis of $U_I$.
The group $G=SU(2)\times SU(2)$ is the double cover of $SO(4)$; it is also known as
the spin group $\text{Spin}(4)$. Any of its elements $g$ can be written
as a couple $(g_L,g_R)$ of two $SU(2)$ group elements. Its Unitary Irreducible Representations (UIR) are labelled
by a couple of (integers or half-integers) spins $(I,J)$: they are finite dimensional and 
the vector space $U_{IJ}=U_I\otimes U_J$ of the representation $(I,J)$ is the tensor product of 
the two $SU(2)$ representations vector spaces $U_I$ and $U_J$. Therefore, the family of vectors
$(\vert I,i \ra \otimes \vert J,j \ra)_{IJij}$ 
form an orthonormal basis of $U_{IJ}$. The action of $g\in G$ in this basis is simply given by:
\be
R^{IJ}(g) \vert I,i \ra \otimes \vert J,j \ra \, = \, R^{IJ}(g_L,g_R)  \vert I,i \ra \otimes \vert J,j \ra \, = \, 
D^I(g_L) \vert I,i \ra \otimes D^J(g_R) \vert J,j \ra \;.
\ee
The $SU(2)$ matrix elements $\langle I,i \vert D^I \vert I,j \rangle$ are the Wigner functions.

\medskip

The space $U_{IJ}$ admits another natural basis which will be useful in the sequel.
This other basis is constructed from the remark that the vector space $U_{IJ}$ decomposes into $SU(2)$
UIR vector spaces $U_K$ as follows:
\be\label{decomposition}
U_{IJ} \; \simeq \; \bigoplus_{K=\vert I-J \vert}^{I+J} U_K \;.
\ee
This decomposition provides indeed another orthonormal basis of $U_{IJ}$, given by the family of vectors
$(\vert K,k\ra)_{Kk}$ where $K \in [\vert I-J \vert,I+J]$ and $k \in [-K,K]$ as usual.
The changing basis formulae are given in terms of the Clebsch-Gordan coefficients $\la Kk \vert IiJj\ra$ as follows:
\ba
\vert Ii \ra \otimes \vert Jj\ra = \sum_{K,k} \la Kk\vert IiJi\ra \, \vert Kk \ra \;\;\;\text{and}\;\;\;
\vert Kk \ra  =  \sum_{IJij}\la Kk\vert IiJi\ra \,\vert Ii \ra \otimes \vert Jj\ra \;.
\ea
To write the action of $G$ on the basis elements $\vert Kk\ra$, it is convenient to find the subgroup 
$H\subset G$ which leaves the subspaces $U_K$ of the decomposition (\ref{decomposition}) invariant and then 
to identify $G$ with the space $G\simeq H\times (G/H)$. In fact, it is immediate to see that $H \simeq SU(2)$,
the coset $G/H$ is isomorphic to the sphere $S^3$ and then we identify $G$ with $SU(2) \times S^3$. Notice that
the identification we have just mentionned is not canonical because $G$ admits many $SU(2)$ subgroups;
therefore, to make this identification well defined, one has to precise which $SU(2)$ subgroup one is talking about.
In our case, the $SU(2)$ subgroup is the diagonal one, i.e. it is the group of the elements $(g_L,g_R)$ 
where $g_L=g_R$. As a result, the explicit mapping between $G$ and $SU(2)\times S^3$ is:
\be\label{isomorphism}
G \longrightarrow SU(2)\times S^3 \;\;\;\;\;\; (g_L,g_R)=(u,ux) \longmapsto (u,x)=(g_L,g_L^{-1}g_R)\, .
\ee
This mapping is of course invertible and its inverse is trivially given by:
\be
SU(2)\times S^3 \longrightarrow G \;\;\;\;\;\; (u,x) \longmapsto (u,ux) \,.
\label{isomorphism2}
\ee
The multiplication law $(g_L,g_R)(g'_L,g'_R)=(g_Lg_L',g_Rg_R')$ induces the multiplication rule
\be(u,x)(u',x')\; = \; (uu',{u'}{}^{-1} x u' x')
\ee
in the $SU(2)\times S^3$ representation of $G$.
In particular, the inverse of the element $(u,x)$ is given by $(u,x)^{-1}=({u}^{-1}, u x^{-1} u^{-1})$.
The diagonal terms $u\equiv (u,1)$ and the pure spherical terms $x \equiv (1,x)$ will be relevant in the following construction.

Let us now come back to the action of $G$ on the the vectors $\vert K,k\rangle$ of the vector space $U_{IJ}$;
this action is best written and simpler using the factorization $SU(2) \times S^3$ of $G$.
Indeed, a simple calculation shows that
\begin{eqnarray}
&&R^{IJ}_{KkL\ell}(u) =R^{IJ}_{KkL\ell}(u,u)\!=\!\sum_{m_1,m_2}\!\la Kk\vert I m_1 J m_2  \ra D^I_{m_1,n_1}(g_L)D^J_{m_2,n_2}(u) \la I n_1 J n_2\vert L\ell \ra\!= \!\delta_{K,L} D^K_{k\ell}(u)\nonumber \\
&&R^{IJ}_{KkL\ell}(x)=R^{IJ}_{KkL\ell}(1,x)= \sum_{ijj'} \la Kk\vert IiJj \ra \, \la Ii Jj'\vert L\ell  \ra \, D^J_{jj'}(x)\;.\label{matrix elements ema}
\end{eqnarray}
where we have introduced the notation $R^{IJ}_{KkL\ell}(g)\equiv \la Kk \vert R^{IJ}(g)\vert L\ell\ra$
for the $SU(2)\times SU(2)$ matrix elements. As expected, we see that $u\in SU(2)$ leaves any
$SU(2)$ representation spaces $U_K$ of the decomposition (\ref{decomposition}) invariant whereas 
$x$ moves the vectors from one $SU(2)$ representation space to another. 
This closes  the brief review on $SU(2)\times SU(2)$ representations theory.

\subsubsection{The vertex amplitude as an integral over several copies of $S^3$}
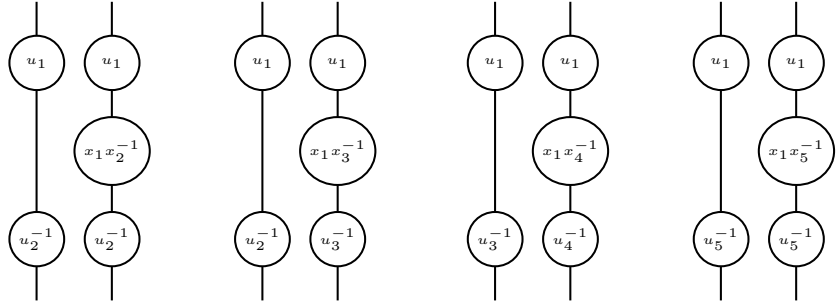
\begin{figure}[h]
\begin{center}
\ifx\JPicScale\undefined\def\JPicScale{0.9}\fi
\psset{unit=\JPicScale mm}
\psset{linewidth=0.3,dotsep=1,hatchwidth=0.3,hatchsep=1.5,shadowsize=1,dimen=middle}
\psset{dotsize=0.7 2.5,dotscale=1 1,fillcolor=black}
\psset{arrowsize=1 2,arrowlength=1,arrowinset=0.25,tbarsize=0.7 5,bracketlength=0.15,rbracketlength=0.15}
\tiny
\begin{pspicture}(0,0)(121.5,44)
\psline(5,44)(5,39)
\psline(5,31)(5,13)
\rput{0}(5,9){\psellipse[](0,0)(4,-4)}
\psline(5,5)(5,0)
\rput(5,35){$u_1$}
\rput(5,9){$u_2^{-1}$}
\rput{0}(5,35){\psellipse[](0,0)(4,-4)}
\psline(16,44)(16,39)
\psline(16,31)(16,27)
\rput{0}(16,9){\psellipse[](0,0)(4,-4)}
\psline(16,5)(16,0)
\rput(16,35){$u_1$}
\rput(16,9){$u_2^{-1}$}
\rput{0}(16,35){\psellipse[](0,0)(4,-4)}
\rput{0}(16,22){\psellipse[](0,0)(5.5,-5)}
\psline(16,17)(16,13)
\rput(16,22){$x_1x_2^{-1}$}
\psline(38,44)(38,39)
\psline(38,31)(38,13)
\rput{0}(38,9){\psellipse[](0,0)(4,-4)}
\psline(38,5)(38,0)
\rput(38,35){$u_1$}
\rput(38,9){$u_2^{-1}$}
\rput{0}(38,35){\psellipse[](0,0)(4,-4)}
\psline(49,44)(49,39)
\psline(49,31)(49,27)
\rput{0}(49,9){\psellipse[](0,0)(4,-4)}
\psline(49,5)(49,0)
\rput(49,35){$u_1$}
\rput(49,9){$u_3^{-1}$}
\rput{0}(49,35){\psellipse[](0,0)(4,-4)}
\rput{0}(49,22){\psellipse[](0,0)(5.5,-5)}
\psline(49,17)(49,13)
\rput(49,22){$x_1x_3^{-1}$}
\psline(72,44)(72,39)
\psline(72,31)(72,13)
\rput{0}(72,9){\psellipse[](0,0)(4,-4)}
\psline(72,5)(72,0)
\rput(72,35){$u_1$}
\rput(72,9){$u_3^{-1}$}
\rput{0}(72,35){\psellipse[](0,0)(4,-4)}
\psline(83,44)(83,39)
\psline(83,31)(83,27)
\rput{0}(83,9){\psellipse[](0,0)(4,-4)}
\psline(83,5)(83,0)
\rput(83,35){$u_1$}
\rput(83,9){$u_4^{-1}$}
\rput{0}(83,35){\psellipse[](0,0)(4,-4)}
\rput{0}(83,22){\psellipse[](0,0)(5.5,-5)}
\psline(83,17)(83,13)
\rput(83,22){$x_1x_4^{-1}$}
\psline(105,44)(105,39)
\psline(105,31)(105,13)
\rput{0}(105,9){\psellipse[](0,0)(4,-4)}
\psline(105,5)(105,0)
\rput(105,35){$u_1$}
\rput(105,9){$u_5^{-1}$}
\rput{0}(105,35){\psellipse[](0,0)(4,-4)}
\psline(116,44)(116,39)
\psline(116,31)(116,27)
\rput{0}(116,9){\psellipse[](0,0)(4,-4)}
\psline(116,5)(116,0)
\rput(116,35){$u_1$}
\rput(116,9){$u_5^{-1}$}
\rput{0}(116,35){\psellipse[](0,0)(4,-4)}
\rput{0}(116,22){\psellipse[](0,0)(5.5,-5)}
\psline(116,17)(116,13)
\rput(116,22){$ x_1x_5^{-1}$}
\end{pspicture}
\end{center}
\caption{Structure of the node $i=1$. Four pairs of edges are attached at each node of the graph: each edge are
colored with a $SU(2)$ representation. The bullets illustrate the inclusions of $SU(2)$ variables $u_i$ or
$S^3$ variables $x_ix_j^{-1}$. Notice that, in the $SU(2)\times SU(2)$ formulation, each pair of lines is associated
to the element $(g_L,g_R)$, $g_L$ corresponding to the left line and $g_R$ to the right one.}
\label{snode}
\end{figure}

We make use of the basic properties on representations theory recalled above to write the 
general formula of the vertex amplitude (\ref{general formula integral}) in the form (\ref{general vertex}).
To do so, one splits the integrations over the group variables $g_i\in G$ in the formula (\ref{general formula integral})
into integrations over the $x_i \in S^3$ variables and integrations over the $u_i \in SU(2)$  variables using the
isomorphism  (\ref{isomorphism}) and one obtains:
\be
V(I_{ij},\omega_i) \; = \; \int (\prod_{i=1}^5 dx_i) (\prod_{i=1}^5 du_i) \, (\otimes_{i=1}^5 v_i) \cdot 
(\bigotimes_{i<j} R^{I_{ij}}(u_i) R^{I_{ij}}(x_ix_j^{-1}) R^{I_{ij}}(u^{-1}_j)  )
\label{Vema}
\ee 
where $R^I(u)\equiv R^I(u,1)$ (resp. $R^I(u,u)$) and $R^I(x)\equiv R^I(1,x)$ (resp. $R^I(1,x)$)
are the matrices of $SU(2)\times SU(2)$ representations $I$ in the $SU(2)\times S^3$ (resp. $SU(2)\times SU(2)$)
formulations. To have a ``geometrical" intuition of this formula, we give a graphical
representation of the integrand in the figure (\ref{general graph}) below.
\begin{figure}
\begin{center}
\ifx\JPicScale\undefined\def\JPicScale{0.9}\fi
\psset{unit=\JPicScale mm}
\psset{linewidth=0.3,dotsep=1,hatchwidth=0.3,hatchsep=1.5,shadowsize=1,dimen=middle}
\psset{dotsize=0.7 2.5,dotscale=1 1,fillcolor=black}
\psset{arrowsize=1 2,arrowlength=1,arrowinset=0.25,tbarsize=0.7 5,bracketlength=0.15,rbracketlength=0.15}
\begin{pspicture}(0,0)(122,119)
\psecurve(45,109)(45,109)(35.3,87.67)(7,81)(7,81)(7,81)
\psecurve(48,109)(48,109)(37.9,86.33)(7,78)(7,78)(7,78)
\psecurve(55,109)(55,109)(49.6,65)(25,17)(25,17)
\psecurve(57,109)(57,109)(52.2,63.67)(27,15)(27,15)
\psecurve(63,109)(63,109)(67.8,63.67)(93,15)(93,15)(93,15)
\psecurve(65,109)(65,109)(70.4,65)(95,17)(95,17)(95,17)
\psecurve(72,109)(72,109)(82.1,86.33)(113,78)(113,78)(113,78)
\psecurve(75,109)(75,109)(84.7,87.67)(113,81)(113,81)(113,81)
\psline[border=0.3](7,73)(113,73)
\psline[border=0.3](7,70)(113,70)
\psecurve(18,21)(18,21)(24.9,42.33)(7,52)(7,52)(7,52)
\psecurve(21,20)(21,20)(27.5,43.67)(7,55)(7,55)(7,55)
\psecurve[border=0.3,curvature=1.0 0.1 0.0](7,64)(7,64)(53.5,51.67)(89,12)(89,12)(89,12)
\psecurve[border=0.3,curvature=1.0 0.1 0.0](7,61)(7,61)(53.5,47.67)(86,11)(86,11)(86,11)
\psecurve[border=0.3,curvature=1.0 0.1 0.0](113,64)(113,64)(66.5,51.67)(31,12)(31,12)(31,12)
\psecurve[border=0.3,curvature=1.0 0.1 0.0](113,61)(113,61)(66.5,47.67)(34,11)(34,11)(34,11)
\psecurve(113,55)(113,55)(92.5,43.67)(98,20)(98,20)(98,20)
\psecurve(113,52)(113,52)(95.1,42.33)(101,21)(101,21)(101,21)
\psecurve(38,7)(38,7)(60,26.33)(80,7)(80,7)(80,7)
\psecurve(40,6)(40,6)(60,23.67)(78,6)(78,6)(78,6)
\rput{90}(39.85,88.33){\psellipse[fillcolor=darkgray,fillstyle=solid](0,0)(2,-1.95)}
\rput{90}(43.1,102.34){\psellipse[fillcolor=gray,fillstyle=solid](0,0)(1.34,1.3)}
\rput{90}(45.7,99.66){\psellipse[fillcolor=gray,fillstyle=solid](0,0)(1.34,1.3)}
\rput{90}(28.8,81){\psellipse[fillcolor=gray,fillstyle=solid](0,0)(1.33,1.3)}
\rput{90}(26.2,83.66){\psellipse[fillcolor=gray,fillstyle=solid](0,0)(1.34,1.3)}
\rput(40.5,105){{\tiny $u_1$}}
\rput(49.6,101){{\tiny $u_1$}}
\rput{90}(53.5,91.66){\psellipse[fillcolor=gray,fillstyle=solid](0,0)(1.33,1.3)}
\rput{90}(56.1,89){\psellipse[fillcolor=gray,fillstyle=solid](0,0)(1.33,1.3)}
\rput{90}(51.55,61.67){\psellipse[fillcolor=darkgray,fillstyle=solid](0,0)(2,-1.95)}
\rput{90}(37.3,31.66){\psellipse[fillcolor=gray,fillstyle=solid](0,0)(1.34,1.3)}
\rput{90}(33.3,31.66){\psellipse[fillcolor=gray,fillstyle=solid](0,0)(1.34,1.3)}
\rput{90}(63.9,89){\psellipse[fillcolor=gray,fillstyle=solid](0,0)(1.33,1.3)}
\rput{90}(66.5,91.66){\psellipse[fillcolor=gray,fillstyle=solid](0,0)(1.33,1.3)}
\rput{90}(74.3,99.66){\psellipse[fillcolor=gray,fillstyle=solid](0,0)(1.33,1.3)}
\rput{90}(76.9,102.34){\psellipse[fillcolor=gray,fillstyle=solid](0,0)(1.34,1.3)}
\rput{90}(82.1,91){\psellipse[fillcolor=darkgray,fillstyle=solid](0,0)(2,-1.95)}
\rput{90}(91.2,81){\psellipse[fillcolor=gray,fillstyle=solid](0,0)(1.33,1.3)}
\rput{90}(93.8,83.66){\psellipse[fillcolor=gray,fillstyle=solid](0,0)(1.33,1.3)}
\rput{90}(71.7,62.33){\psellipse[fillcolor=darkgray,fillstyle=solid](0,0)(2,-1.95)}
\rput(50.9,95.67){{\tiny $u_1$}}
\rput(58.7,85){{\tiny $u_1$}}
\rput(62.6,85){{\tiny $u_1$}}
\rput(69.1,95.67){{\tiny $u_1$}}
\rput(70.4,101){{\tiny $u_1$}}
\rput(79.5,105){{\tiny $u_1$}}
\rput(34,78.33){{\tiny $u_2^{-1}$}}
\rput(22.3,87.67){{\tiny $u_2^{-1}$}}
\rput(44.4,83.67){{\tiny$x_1x_2^{-1}$}}
\rput{90}(21,73){\psellipse[fillcolor=gray,fillstyle=solid](0,0)(1.33,1.3)}
\rput{90}(23.6,70.34){\psellipse[fillcolor=gray,fillstyle=solid](0,0)(1.34,1.3)}
\rput{90}(21,62.34){\psellipse[fillcolor=gray,fillstyle=solid](0,0)(1.34,1.3)}
\rput{90}(19.7,58.34){\psellipse[fillcolor=gray,fillstyle=solid](0,0)(1.34,1.3)}
\rput{90}(15.8,53){\psellipse[fillcolor=gray,fillstyle=solid](0,0)(1.33,1.3)}
\rput{90}(13.2,50.34){\psellipse[fillcolor=gray,fillstyle=solid](0,0)(1.34,1.3)}
\rput(27.5,67.67){{\tiny $u_2$}}
\rput(17.1,75.67){{\tiny $u_2$}}
\rput(24.9,63.67){{\tiny $u_2$}}
\rput(23.6,55.67){{\tiny $u_2$}}
\rput(19.7,53){{\tiny $u_2$}}
\rput(15.8,46.33){{\tiny $u_2$}}
\rput{90}(51.55,48.33){\psellipse[fillcolor=darkgray,fillstyle=solid](0,0)(2,-1.95)}
\rput{90}(67.8,47.67){\psellipse[fillcolor=darkgray,fillstyle=solid](0,0)(2,-1.95)}
\rput{90}(24.25,41){\psellipse[fillcolor=darkgray,fillstyle=solid](0,0)(2,-1.95)}
\rput{90}(60,23){\psellipse[fillcolor=darkgray,fillstyle=solid](0,0)(2,-1.95)}
\rput{90}(96.4,43.67){\psellipse[fillcolor=darkgray,fillstyle=solid](0,0)(2,-1.95)}
\rput{90}(60,69.67){\psellipse[fillcolor=darkgray,fillstyle=solid](0,0)(2,-1.95)}
\rput{90}(21,26.34){\psellipse[fillcolor=gray,fillstyle=solid](0,0)(1.34,1.3)}
\rput{90}(24.9,27.66){\psellipse[fillcolor=gray,fillstyle=solid](0,0)(1.34,1.3)}
\rput{90}(43,28.34){\psellipse[fillcolor=gray,fillstyle=solid](0,0)(1.33,1.3)}
\rput{90}(44.7,25){\psellipse[fillcolor=gray,fillstyle=solid](0,0)(1.33,1.3)}
\rput{90}(43.1,14.34){\psellipse[fillcolor=gray,fillstyle=solid](0,0)(1.34,1.3)}
\rput{90}(43.1,10.34){\psellipse[fillcolor=gray,fillstyle=solid](0,0)(1.34,1.3)}
\rput{90}(75.3,25.34){\psellipse[fillcolor=gray,fillstyle=solid](0,0)(1.33,1.3)}
\rput{90}(77,28.33){\psellipse[fillcolor=gray,fillstyle=solid](0,0)(1.33,1.3)}
\rput{90}(82.7,32){\psellipse[fillcolor=gray,fillstyle=solid](0,0)(1.33,1.3)}
\rput{90}(87,31.34){\psellipse[fillcolor=gray,fillstyle=solid](0,0)(1.33,1.3)}
\rput{90}(94,28){\psellipse[fillcolor=gray,fillstyle=solid](0,0)(1.33,1.3)}
\rput{90}(98.3,26.34){\psellipse[fillcolor=gray,fillstyle=solid](0,0)(1.33,1.3)}
\rput{90}(75.6,14.34){\psellipse[fillcolor=gray,fillstyle=solid](0,0)(1.34,1.3)}
\rput{90}(75.6,10.34){\psellipse[fillcolor=gray,fillstyle=solid](0,0)(1.34,1.3)}
\rput{90}(97.7,73){\psellipse[fillcolor=gray,fillstyle=solid](0,0)(1.33,1.3)}
\rput{90}(95.1,70.34){\psellipse[fillcolor=gray,fillstyle=solid](0,0)(1.34,1.3)}
\rput{90}(99,62.34){\psellipse[fillcolor=gray,fillstyle=solid](0,0)(1.33,1.3)}
\rput{90}(100.3,58.34){\psellipse[fillcolor=gray,fillstyle=solid](0,0)(1.33,1.3)}
\rput{90}(104.2,53){\psellipse[fillcolor=gray,fillstyle=solid](0,0)(1.33,1.3)}
\rput{90}(106.8,50.34){\psellipse[fillcolor=gray,fillstyle=solid](0,0)(1.33,1.3)}
\rput(32,35){{\tiny$u_3^{-1}$}}
\rput(42,34){{\tiny$u_3^{-1}$}}
\rput(16,29){{\tiny$u_3^{-1}$}}
\rput(26,24){{\tiny$u_3^{-1}$}}
\rput(40,27){{\tiny$u_3$}}
\rput(45,21){{\tiny$u_3$}}
\rput(48,8){{\tiny$u_3$}}
\rput(39.2,13){{\tiny$u_3$}}
\rput(73,7){{\tiny $u_4^{-1}$}}
\rput(76,18){{\tiny $u_4^{-1}$}}
\rput(70,26){{\tiny $u_4^{-1}$}}
\rput(78,34){{\tiny $u_4^{-1}$}}
\rput(82,27){{\tiny $u_4^{-1}$}}
\rput(55,58){{\tiny $x_1 x_3^{-1}$}}
\rput(89,35){{\tiny $u_4^{-1}$}}
\rput(102,24){{\tiny $u_4$}}
\rput(93,25){{\tiny $u_4$}}
\rput(79,62){{\tiny $x_1 x_4^{-1}$}}
\rput(89.9,91.67){{\tiny $x_1 x_5^{-1}$}}
\rput(87.3,91.67){}
\rput(60,65){{\tiny $x_2 x_5^{-1}$}}
\rput(48.3,43.67){{\tiny $x_2 x_4^{-1}$}}
\rput(18.4,39.67){{\tiny $x_2 x_3^{-1}$}}
\rput(60,18.33){{\tiny $x_3 x_4^{-1}$}}
\rput(70.4,43.67){{\tiny $x_3 x_5^{-1}$}}
\rput(101.6,39.67){{\tiny $x_4 x_5^{-1}$}}
\rput(111,49){{\tiny $u_5^{-1}$}}
\rput(108,56){{\tiny $u_5^{-1}$}}
\rput(97,55){{\tiny $u_5^{-1}$}}
\rput(95,64){{\tiny $u_5^{-1}$}}
\rput(92,67){{\tiny $u_5^{-1}$}}
\rput(101,76){{\tiny $u_5^{-1}$}}
\rput(87,79){{\tiny $u_5^{-1}$}}
\rput(92,87){{\tiny $u_5^{-1}$}}
\pspolygon[](45,119)(75,119)(75,115)(45,115)
\psline(45,115)(45,113)
\psline(48,115)(48,113)
\psline(55,115)(55,113)
\psline(57,115)(57,113)
\psline(63,115)(63,113)
\psline(65,115)(65,113)
\psline(72,115)(72,113)
\psline(75,115)(75,113)
\rput(59,117){$v_1$}
\pspolygon[](118,81)(122,81)(122,52)(118,52)
\psline(116,81)(118,81)
\psline(116,78)(118,78)
\psline(116,73)(118,73)
\psline(116,70)(118,70)
\psline(116,64)(118,64)
\psline(116,61)(118,61)
\psline(116,55)(118,55)
\psline(116,52)(118,52)
\rput(120,67){$v_5$}
\psline(2,81)(4,81)
\psline(2,78)(4,78)
\psline(2,73)(4,73)
\psline(2,70)(4,70)
\psline(2,64)(4,64)
\psline(2,61)(4,61)
\psline(2,55)(4,55)
\psline(2,52)(4,52)
\rput(0,67){$v_2$}
\pspolygon[](15,18)(36.33,2.49)(34,-1)(12.67,14.51)
\psline(16,19)(15,18)
\psline(18,18)(17,17)
\psline(23,14)(22,13)
\psline(25,13)(24,12)
\psline(29,10)(28,9)
\psline(31,8)(30,7)
\psline(36,5)(35,4)
\psline(38,4)(36,2)
\rput(24,9){$v_3$}
\pspolygon[](-2,81)(2,81)(2,52)(-2,52)
\pspolygon[](106.28,14.34)(84.8,-0.95)(82.23,2.37)(103.72,17.66)
\psline(88,8)(89,7)
\psline(90,10)(91,9)
\psline(94,13)(95,12)
\psline(97,15)(98,14)
\psline(102,18)(103,17)
\psline(100,17)(101,16)
\psline(81,4)(82,3)
\psline(83,5)(84,4)
\rput(94,8){$v_4$}
\end{pspicture}
\caption{This picture is a graphical representation of the integrand in the formula (\ref{Vema}) 
defining the vertex amplitude. Each line are doubled because it carries a representation  of 
$SU(2)\times SU(2)$ and the single lines in the pair colored with $(I,J)$ are colored by $I$ and $J$
separately. Furthermore, the single lines are endowed with bullets that represent the insertion of $SU(2)$ 
group elements: the small ones are associated to diagonal elements $u_i \in SU(2)$ whereas the big ones
are associated to spherical elements $x_ix_j^{-1}\in S^3$.
The vectors $v_i$ are represented by boxes and they are contracted with the free ends of the graph.}

\label{general graph}
\end{center}
\end{figure}
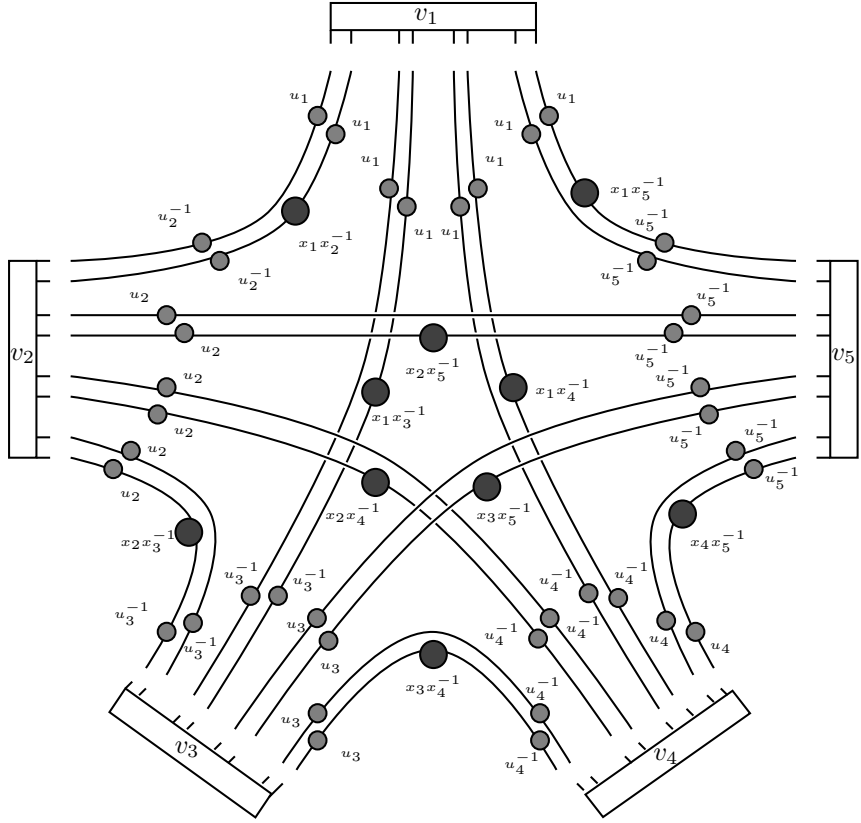
In the models we are going to consider explicitely in the sequel, we can  perform the integrations over the $u_i$
variables; therefore we formally perform the integration over the $u_i$'s in the general formula (\ref{Vema}) and we 
obtain a formula for the vertex amplitude as an integral over 5 copies of $S^3$ only:
\ba\label{integral expression}
V(I_{ij},\omega_i) \; = \; \int (\prod_{i=1}^5 dx_i) \, (\otimes_{i=1}^5 \nu_i) \cdot 
(\bigotimes_{i<j} R^{I_{ij}}(x_ix_j^{-1})) \,.
\ea
The integrations over the five $SU(2)$ variables $u_i$ have been hidden in the following definition of
the vectors $\nu_i \in \otimes_{j\neq i} U_{I_{ij}}^*$:
\ba\label{Vi}
\nu_i \; \equiv \;  \sum_{(a_{ij})} \alpha_i^{(a_{ij})} \,
\int du \; (\otimes_{j>i} \langle v_{a_{ij}} \vert R^{I_{ij}}(u_i) )\, \otimes \,
(\otimes_{j<i} R^{I_{ji}}(u){}^{-1}\vert v_{a_{ij}}\rangle)
\ea
where we have used the explicit decomposition of the vectors $v_i \in \otimes_{j\neq i} U_{I_{ij}}^*$ 
given in the introductive part of Section 1.2. This formula will be much more explicit when we consider the particular
Spin-Foam models we are interested in. For the moment, for pedagogical purposes,
we propose a picturial representation in the figure (\ref{snode}) of the argument of 
the previous integral (\ref{Vi}) when $i=1$.

Before considering specific examples, let us add one more important remark. The vertex amplitude can be trivially
reformulated as an integral over 10 copies of $G$ as follows:
\ba\label{integralV}
V(I_{ij},\omega_i) \; = \; \int (\prod_{i<j} dx_{ij}) \, C(x_{ij}) \, ( \otimes_{i=1}^5 \nu_i ) \cdot 
(\bigotimes_{i<j} R^{I_{ij}}(x_{ij})) 
\ea
where the contraint $C(x_{ij})$ is a distribution which
imposes, rougthly speaking, $x_{ij}$ to be a ``coboundary", i.e. of the form $x_ix_j^{-1}$. An
explicit formula for $C(x_{ij})$ is simply given by the integral:
\be
C(x_{ij}) \; = \; \int (\prod_{i=1}^5 dx_i) \, \prod_{i\neq j} \delta(x_{ij}^{-1}x_ix_j^{-1}) 
\ee
where $\delta$ is the $SU(2)$ delta distribution. It is possible to perform the above integration
whose result is simply given by the product of five delta distributions:
\be\label{definition of C}
C(x_{ij}) \; = \; \delta(x_{123}) \, \delta(x_{234}) \, \delta(x_{345}) \, \delta(x_{451}) \, \delta(x_{512})
\ee
where $x_{ijk}=x_{ij}x_{jk}x_{ki}$ and, by convention, $x_{ij}=x_{ji}^{-1}$. The interpretation
of the constraint $C(x_{ij})$  will become clear  
in the last Section where we make the link with the canonical quantization.
To conclude, we underline that we have finally found the desired formula (\ref{general vertex}) for 
the vertex amplitude with the anounced expression of the distribution $C(x_{ij})$
and  the model dependent function ${\cal V}(I_{ij},\omega_i;x_{ij})=( \otimes_{i=1}^5 \nu_i ) \cdot 
(\bigotimes_{i<j} R^{I_{ij}}(x_{ij}))$ is a particular contraction of five $SU(2)$ matrices.

\subsection{Vertices of particular models}
\begin{figure}[h]
\psfrag{+}{$\iota_+(\alpha)$}
\psfrag{-}{$\iota_-(\alpha)$}
\psfrag{0}{$\iota_0(\alpha)$}
\centering
\includegraphics[scale=0.8]{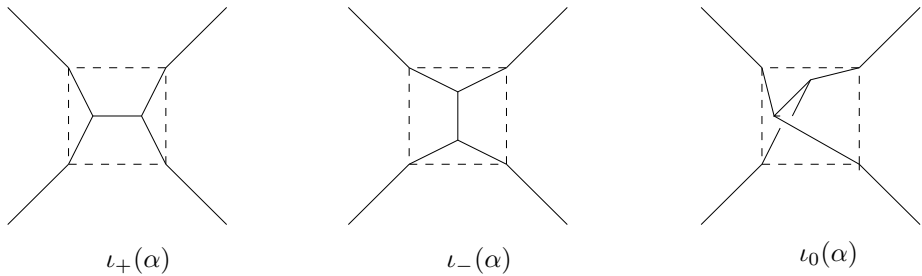}
\caption{\small{The three canonical basis of the space of 4-valent intertwiners. The intermediate channel
is endowed with the representation $\alpha$.}}
\label{intertwiners basis}
\end{figure}

This part is devoted to study some aspects of the vertex amplitude (\ref{integral expression}) 
for the topological model, the BC model and the EPR model. 
In fact, these models differ only by the choice of the intertwiners $\omega_i$ or equivalently the vectors $v_i$
which are their building blocks.
Thus, to understand the construction of these models and their differences, one has to understand the definition
of their associated intertwiners. For that purpose, let us start by recalling basic properties of intertwiners.
First of all, in Spin-Foam models, we are interested in 4-valent intertwiners only. The 4-valent intertwiners
between four given representations form a (normed) vector space of finite dimension. In the case
where $G=SU(2)$, one can exhibit three 
canonical (natural) orthogonal basis 
(labelled by an index $\epsilon \in \{+,-,0\}$ that indicates the ``coupling channel") presented in the figure (\ref{intertwiners basis}). 
Whatever the basis we choose, any of its element is
completely characterized by the  representation appearing in the intermediate channel in the tensor
product decomposition. Therefore, one often identifies the element of each basis with a representation.
We will use the notations $\iota_\epsilon(\alpha)$ to denote the $SU(2)$ intertwiner in the basis $\epsilon$ with
intermediate representation $\alpha$. One can make used of these results to construct basis of $SU(2)\times SU(2)$
4-valent intertwiners. In particular, one can naturally exhibit nine ``tensor product" basis labelled by a couple
$(\epsilon,\epsilon')$.  
However, we will consider in the sequel only the three basis of the type
$(\epsilon,\epsilon)$ which will be labelled by a single $\epsilon$ for simplicity: elements of the basis
$\epsilon$ are denoted $\iota_\epsilon(\alpha)$ as in the $SU(2)$ case but with the difference that $\alpha$
is now a couple of $SU(2)$ representations.

Now, we are ready to define the intertwiner $\omega_i$ for the model we are interested in.
Afterwards, we are going to make the general abstract formula of the vertex amplitude more concrete and
more useful for studying its properties.

\subsubsection{The topological model}
\begin{figure}[h]
\psfrag{o1}{$\omega_1$}
\psfrag{o2}{$\omega_2$}
\psfrag{o3}{$\omega_3$}
\psfrag{o4}{$\omega_4$}
\psfrag{o5}{$\omega_5$}
\psfrag{I12}{$I_{12}$}
\psfrag{I13}{$I_{13}$}
\psfrag{I14}{$I_{14}$}
\psfrag{I15}{$I_{15}$}
\psfrag{I23}{$I_{23}$}
\psfrag{I24}{$I_{24}$}
\psfrag{I25}{$I_{25}$}
\psfrag{I34}{$I_{34}$}
\psfrag{I35}{$I_{35}$}
\psfrag{I45}{$I_{45}$}
\centering
\includegraphics[scale=0.6]{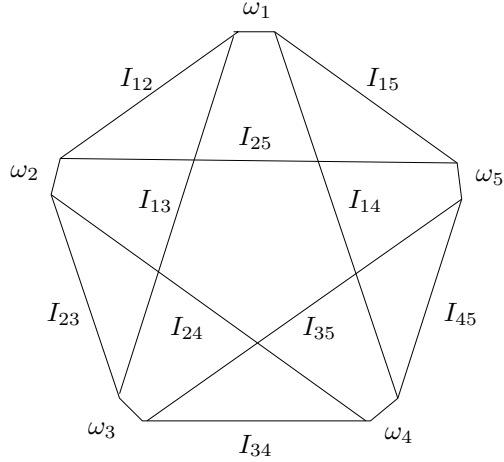}
\caption{\small{Pictorial representation of a 15j symbol: vertices are labelled by representations $\omega_i$
and edges by representations $I_{ij}$.}}
\label{15}
\end{figure}

We start with the simplest, certainly the more mathematically precise but non-physical model.
The topological model is closely related to $BF$ theory with gauge group $SU(2)$. 
More precisely, given a triangulation $\cal T$ of a 4-dimensional 
manifold $\cal M$, one can discretize the $BF$ action to be well-defined on this triangulation and the path
integral ${\cal Z}_{BF}({\cal T})$ 
of the discretized action can be formulated as a state sum or equivalently a Spin-Foam model:
\ba
{\cal Z}_{BF}({\cal T}) \; = \; \sum_{\{j_f\},\{\omega_t\}} \prod_{f \in {\cal T}^2} \text{dim}(j_f) \,
\prod_{t \in {\cal T}^3} \text{dim}(\omega_t)^{-1} \, \prod_{s \in {\cal T}^4}V_{BF}(\omega_{t_s},j_{f_s})
\ea
where we have used notations of (\ref{generalSFamp}); 
we have identified the intertwiners $\omega_t$ with the 
associated representation and $V_{BF}$ is the vertex amplitude completely defined by the graph (\ref{15}).
This amplitude is known as a 15j symbol and can be formulated as a finite sum of products of 6j symbols
\ba
V_{BF}(I_{ij},\omega_i) & = & \sum_K \frac{1}{d_{\omega_1}d_{\omega_5} d_K^2}
\left\{\begin{array}{ccc}
\omega_1 & I_{12} & I_{13}\\
\omega_2 & I_{25} & K
\end{array}\right\}
\left\{\begin{array}{ccc}
\omega_2 & \omega_3 & I_{13}\\
I_{23} & I_{13} & K
\end{array}\right\} \nonumber \\
&& \left\{\begin{array}{ccc}
I_{35} & I_{24} & I_{34}\\
\omega_3 & \omega_4 & K
\end{array}\right\}
\left\{\begin{array}{ccc}
\omega_1 & I_{14} & I_{15}\\
I_{25} & \omega_5 & K
\end{array}\right\}
\left\{\begin{array}{ccc}
I_{45} & \omega_5 & I_{34}\\
I_{14} & \omega_4 & K
\end{array}\right\}.
\ea
The 6j symbols are the totally symmetrized 6j symbols defined, for example, in the chapter 6 
of the book \cite{edmonds}.
Note that the sum is finite and then the vertex amplitude is well-defined.
However, the state sum is generally divergent;
it can be made convergent by gauge fixing or by turning classical groups into quantum groups. 
The state sum is a (formal) PL invariant, i.e. invariant under homeomorphisms. 

We have voluntarily not given neither the interwiners $\omega_i^{BF}$ nor the vectors $v_i$ defining the model
according to the previous Section. Indeed, such a formulation is not very useful for the topological model
and the description of the previous Section is naturally  adpated for $SU(2)\times SU(2)$ Spin-Foam
models and not really for $SU(2)$ Spin-Foam models.

\subsubsection{The Barrett-Crane model}
The Barrett-Crane model has been constructed as a step towards the covariant quantization of four dimensional
pure Euclidean or Lorentzian gravity \`a la Plebanski. Here, we consider exclusively the Euclidean case. 
The BC model is then a state sum associated to a triangulation $\cal T$ of a 4-manifold $\cal M$
which is supposed to reproduce the path integral ${\cal Z}_{Pl}({\cal T})$ of a discretized version of the Plebanski 
action. However, the link between the BC model and gravity is somehow misleading. Indeed,
the BC state sum has been constructed heuristically as a modification of the $SU(2)\times SU(2)$ topological 
state sum according to the following rules: 
representations coloring the faces of the 4-simplex are supposed to be simple, i.e. of the form $(I_{ij},I_{ij})$;
the intertwiners $\omega_i^{BC}$ associated to the tetrahedra are also called simple or BC intertwiners we will
recall the definition in the sequel; the vertex amplitude $V_{BC}$ associated to 
the 4-simplices are the so-called 10j symbols
whose definition will also be recalled later. The BC model does not say anything concerning the amplitudes
${\cal A}_2$ and ${\cal A}_3$ associated to the faces and the tetrahedra of the triangulation. However,
many arguments lead to certain expressions of ${\cal A}_2$ and ${\cal A}_3$ and the corresponding state sums 
have been numerically tested \cite{num}. Anyway, we will not consider these amplitudes in this paper.
 
Let us concentrate on the construction of the vertex amplitude $V_{BC}$ whose basic ingredient 
is the simple intertwiner.
A simple $n$-valent intertwiner is such that any of its decompositions into 3-valent intertwiners introduce only
simple representations in the intermediate channel. The simple intertwiner has been studied intensively in the literature;
in particular it was shown to be unique up to a global normalization \cite{unique}. This property makes clear that the vertex 
amplitude
of the BC model is a function $V(I_{ij},\omega_i^{BC})$ of only 10 representations and it is called a 10j symbol.
To precisely define the simple intertwiner $\omega_i^{BC}$, it is more convenient to start with the formula (\ref{omega}) 
which shows that  $\omega_i^{BC}$ is completely determined by the choice of a ``simple" vector 
$v_i^{BC} \in \otimes_{j\neq i} U_{I_{ij}J_{ij}}^*$ where $(I_{ij},J_{ij})$ is a $SU(2)\times SU(2)$ UIR . 
If $(I_{ij},J_{ij})$ is a simple representation, i.e. $I_{ij}=J_{ij}$, then
the associated vector space admits an unique normalized (diagonal) $SU(2)$ invariant vector $w$ (or $\vert w \ra$) 
which
we identify with its dual $\la w \vert \in V_{I_{ij}}^*$. In that case, indeed, the decomposition (\ref{decomposition})
of $U_{I_{ij}J_{ij}}$ into $SU(2)$ representations contains the space $U_0$ which is the one dimensional space
of diagonal $SU(2)$ invariant states.
The simple vector is in fact the tensor product of these invariant vectors: $\omega_i^{BC}=w^{\otimes 4}$. 
As a result, the expression of the simple intertwiner in the tensor product basis reads:
\ba
\omega_i^{BC} \; = \; \frac{1}{\prod_{j\neq i}\sqrt{ d_{I_{ij}}}} \, \sum_{\alpha} d_{\alpha} \, 
\iota_\epsilon(\alpha) 
\ea
where the sum runs over simple representations $\alpha \equiv(\alpha,\alpha)$ 
only and is finite. An important property is that the previous sum is independent on the choice 
of the basis $\epsilon$.
Using this formula of the simple intertwiner, one finds immediately the vertex amplitude of the BC model
\ba
V_{BC}(I_{ij},\omega_i^{BC}) \; = \; \frac{1}{\prod_{i\neq j} d_{I_{ij}}}\,
\sum_\alpha d_\alpha \, V_{BF}(I_{ij},\alpha)^2
\ea
as a sum of BF amplitudes $V_{BF}$ which are $SU(2)$ 15j symbols. The sum runs over 
simple representations only and is independent on the choice of the intertwiners defining the
15j symbol. Such a formula is too cumbersome to be useful and one prefers to use
the integral formulation (\ref{integralV}) of the amplitude to study its physical properties.
This integral formula simplifies indeed drastically because the $SU(2)$ integral defining $\nu_i$ (\ref{Vi}) 
becomes trivial due to the $SU(2)$ invariance of the vectors $v_i$, and reads 
\be
V_{BC}(I_{ij},\omega_i) \; = \; \int (\prod_{i=1}^5 dx_i) \, \la w^{\otimes 10} \vert
\bigotimes_{i<j} R^{I_{ij}}(1,x_ix_j^{-1}) \, \vert w^{\otimes 10} \rangle.
\label{BCema2}
\ee
Using the second equations in \eqref{matrix elements ema}, one obtains the following integral formula
for the 10j symbol:
\ba\label{VBCintegral}
V_{BC}(I_{ij},\omega_i^{BC}) \; = \; 
\int \prod_{i\neq j}  dx_{ij} \, \frac{\chi_{I_{ij}}(x_{ij})}{d_{I_{ij}}}  \, C(x_{ij}) \; = \;
\int \prod_{i=1}^5 dx_i \, \prod_{i<j} \frac{\chi_{I_{ij}}(x_ix_j^{-1})}{d_{I_{ij}}} 
\ea
where $\chi_I(x)$ is the $SU(2)$ character of $x$ in the representation $I$. Up to some normalization
factors, the previous formula coincides with the Euclidean 10j symbols. This integral formulation was very
useful to study the classical behavior of the Euclidean BC model. Let us finish this brief presentation of
the BC model with two important remarks.

\medskip

{\bf Remark 1.}
 The previous calculation can be done in a completely graphical way. Indeed, the ``black" boxes representing the
vectors $v_i^{BC}$ in (\ref{general graph}) reduce to the following form
\be
\ifx\JPicScale\undefined\def\JPicScale{1}\fi
\psset{unit=\JPicScale mm}
\psset{linewidth=0.3,dotsep=1,hatchwidth=0.3,hatchsep=1.5,shadowsize=1,dimen=middle}
\psset{dotsize=0.7 2.5,dotscale=1 1,fillcolor=black}
\psset{arrowsize=1 2,arrowlength=1,arrowinset=0.25,tbarsize=0.7 5,bracketlength=0.15,rbracketlength=0.15}
\begin{pspicture}(0,0)(83,14)
\pspolygon[](2,7.4)(34,7.4)(34,2.6)(2,2.6)
\psline(2,9.8)(2,7.4)
\psline(5.2,9.8)(5.2,7.4)
\psline(11.6,9.8)(11.6,7.4)
\psline(14.8,9.8)(14.8,7.4)
\psline(21.2,9.8)(21.2,7.4)
\psline(24.4,9.8)(24.4,7.4)
\psline(30.8,9.8)(30.8,7.4)
\psline(34,9.8)(34,7.4)
\rput(18,5){{\small$v_i^{BC}$}}
\rput(40,5){=}
\psecurve(45.6,11.6)(45.6,11.6)(47.2,6.8)(48.8,11.6)(48.8,11.6)(48.8,11.6)
\psecurve(56.6,11.6)(56.6,11.6)(58.2,6.8)(59.8,11.6)(59.8,11.6)(59.8,11.6)
\psecurve(67.6,11.6)(67.6,11.6)(69.2,6.8)(70.8,11.6)(70.8,11.6)(70.8,11.6)
\psecurve(78.4,11.8)(78.4,11.8)(80,7)(81.6,11.8)(81.6,11.8)(81.6,11.8)(81.6,11.8)
\psline[linestyle=dotted](47.2,6.8)(47.2,3.2)
\psline[linestyle=dotted](58.2,6.8)(58.2,3.2)
\psline[linestyle=dotted](69.2,6.8)(69.2,3.2)
\psline[linestyle=dotted](80.2,6.8)(80.2,3.2)
\rput(1,14){$I_{ij}$}
\rput(11,14){$I_{ik}$}
\rput(6,14){$I_{ij}$}
\rput(16,14){$I_{ik}$}
\rput(21,14){$I_{il}$}
\rput(25,14){$I_{il}$}
\rput(30,14){$I_{im}$}
\rput(35,14){$I_{im}$}
\rput(45,14){$I_{ij}$}
\rput(56,14){$I_{ik}$}
\rput(50,14){$I_{ij}$}
\rput(61,14){$I_{ik}$}
\rput(67,14){$I_{il}$}
\rput(72,14){$I_{il}$}
\rput(78,14){$I_{im}$}
\rput(83,14){$I_{im}$}
\rput(48,1){$0$}
\rput(59,1){$0$}
\rput(70,1){$0$}
\rput(81,1){$0$}
\end{pspicture}\ee
where the dashed lines represent spin $0$ representation. We see explicitely that $v_i^{BC}$ project into
diagonal $SU(2)$ invariant vectors. Furthermore, the 3j vectors involving a spin 0 representation are proportional
to the ``identity" according to the following pictorial rule
\be
\ifx\JPicScale\undefined\def\JPicScale{1}\fi
\psset{unit=\JPicScale mm}
\psset{linewidth=0.3,dotsep=1,hatchwidth=0.3,hatchsep=1.5,shadowsize=1,dimen=middle}
\psset{dotsize=0.7 2.5,dotscale=1 1,fillcolor=black}
\psset{arrowsize=1 2,arrowlength=1,arrowinset=0.25,tbarsize=0.7 5,bracketlength=0.15,rbracketlength=0.15}
\begin{pspicture}(0,0)(83,14)
\psecurve(30.6,11.6)(30.6,11.6)(32.2,6.8)(33.8,11.6)(33.8,11.6)(33.8,11.6)
\psecurve(45.6,11.6)(45.6,11.6)(47.2,6.8)(48.8,11.6)(48.8,11.6)(48.8,11.6)
\psline[linestyle=dotted](47.2,6.8)(47.2,3.2)
\rput(40,8){=}
\rput(57,9){$\sqrt{d_{I_{ij}}}$.}
\rput(45,14){$I_{ij}$}
\rput(50,14){$I_{ij}$}
\rput(47,0){$0$}
\rput(30,14){$I_{ij}$}
\rput(35,14){$I_{ij}$}
\end{pspicture}\ee

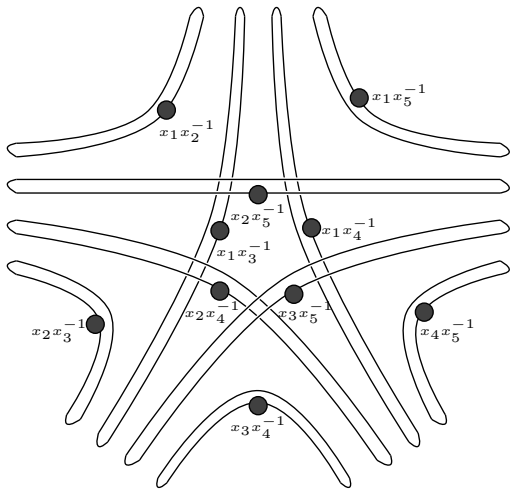
\begin{figure}
\begin{center}
\ifx\JPicScale\undefined\def\JPicScale{0.6}\fi
\psset{unit=\JPicScale mm}
\psset{linewidth=0.3,dotsep=1,hatchwidth=0.3,hatchsep=1.5,shadowsize=1,dimen=middle}
\psset{dotsize=0.7 2.5,dotscale=1 1,fillcolor=black}
\psset{arrowsize=1 2,arrowlength=1,arrowinset=0.25,tbarsize=0.7 5,bracketlength=0.15,rbracketlength=0.15}
\begin{pspicture}(0,0)(115.15,111.15)
\psecurve(45,109)(45,109)(35.3,87.67)(7,81)(7,81)(7,81)
\psecurve(48,109)(48,109)(37.9,86.33)(7,78)(7,78)(7,78)
\psecurve(55,109)(55,109)(49.6,65)(25,17)(25,17)
\psecurve(57,109)(57,109)(52.2,63.67)(27,15)(27,15)
\psecurve(63,109)(63,109)(67.8,63.67)(93,15)(93,15)(93,15)
\psecurve(65,109)(65,109)(70.4,65)(95,17)(95,17)(95,17)
\psecurve(72,109)(72,109)(82.1,86.33)(113,78)(113,78)(113,78)
\psecurve(75,109)(75,109)(84.7,87.67)(113,81)(113,81)(113,81)
\psline[border=0.3](7,73)(113,73)
\psline[border=0.3](7,70)(113,70)
\psecurve(18,21)(18,21)(24.9,42.33)(7,52)(7,52)(7,52)
\psecurve(21,20)(21,20)(27.5,43.67)(7,55)(7,55)(7,55)
\psecurve[border=0.3,curvature=1.0 0.1 0.0](7,64)(7,64)(53.5,51.67)(89,12)(89,12)(89,12)
\psecurve[border=0.3,curvature=1.0 0.1 0.0](7,61)(7,61)(53.5,47.67)(86,11)(86,11)(86,11)
\psecurve[border=0.3,curvature=1.0 0.1 0.0](113,64)(113,64)(66.5,51.67)(31,12)(31,12)(31,12)
\psecurve[border=0.3,curvature=1.0 0.1 0.0](113,61)(113,61)(66.5,47.67)(34,11)(34,11)(34,11)
\psecurve(113,55)(113,55)(92.5,43.67)(98,20)(98,20)(98,20)
\psecurve(113,52)(113,52)(95.1,42.33)(101,21)(101,21)(101,21)
\psecurve(38,7)(38,7)(60,26.33)(80,7)(80,7)(80,7)
\psecurve(40,6)(40,6)(60,23.67)(78,6)(78,6)(78,6)
\rput{90}(39.85,88.33){\psellipse[fillcolor=darkgray,fillstyle=solid](0,0)(2,-1.95)}
\rput{90}(51.55,61.67){\psellipse[fillcolor=darkgray,fillstyle=solid](0,0)(2,-1.95)}
\rput{90}(82.1,91){\psellipse[fillcolor=darkgray,fillstyle=solid](0,0)(2,-1.95)}
\rput{90}(71.7,62.33){\psellipse[fillcolor=darkgray,fillstyle=solid](0,0)(2,-1.95)}
\rput(44.4,83.67){{\tiny$x_1x_2^{-1}$}}
\rput{90}(51.55,48.33){\psellipse[fillcolor=darkgray,fillstyle=solid](0,0)(2,-1.95)}
\rput{90}(67.8,47.67){\psellipse[fillcolor=darkgray,fillstyle=solid](0,0)(2,-1.95)}
\rput{90}(24.25,41){\psellipse[fillcolor=darkgray,fillstyle=solid](0,0)(2,-1.95)}
\rput{90}(60,23){\psellipse[fillcolor=darkgray,fillstyle=solid](0,0)(2,-1.95)}
\rput{90}(96.4,43.67){\psellipse[fillcolor=darkgray,fillstyle=solid](0,0)(2,-1.95)}
\rput{90}(60,69.67){\psellipse[fillcolor=darkgray,fillstyle=solid](0,0)(2,-1.95)}
\rput(57,57){{\tiny $x_1 x_3^{-1}$}}
\rput(80,62){{\tiny $x_1 x_4^{-1}$}}
\rput(91,91.67){{\tiny $x_1 x_5^{-1}$}}
\rput(60,65){{\tiny $x_2 x_5^{-1}$}}
\rput(50.0,43.67){{\tiny $x_2 x_4^{-1}$}}
\rput(16.4,39.67){{\tiny $x_2 x_3^{-1}$}}
\rput(60,18.33){{\tiny $x_3 x_4^{-1}$}}
\rput(70.4,43.67){{\tiny $x_3 x_5^{-1}$}}
\rput(101.6,39.67){{\tiny $x_4 x_5^{-1}$}}
\psecurve(45,109)(45,109)(47,111)(48,109)(48,109)(48,109)
\psecurve(55,109)(55,109)(56,111)(57,109)(57,109)(57,109)(57,109)(57,109)
\psecurve(63,109)(63,109)(64,111)(65,109)(65,109)(65,109)
\psecurve(72,109)(72,109)(73,111)(75,109)(75,109)(75,109)
\psecurve(113,81)(113,81)(115,79)(113,78)(113,78)(113,78)
\psecurve(113,73)(113,73)(115,71)(113,70)(113,70)(113,70)
\psecurve(113,64)(113,64)(115,62)(113,61)(113,61)(113,61)
\psecurve(113,55)(113,55)(115,53)(113,52)(113,52)
\psecurve(7,52)(7,52)(5,54)(7,55)(7,55)(7,55)
\psecurve(7,61)(7,61)(5,63)(7,64)(7,64)(7,64)
\psecurve(7,70)(7,70)(5,72)(7,73)(7,73)(7,73)
\psecurve(7,78)(7,78)(5,80)(7,81)(7,81)(7,81)
\psecurve(18,21)(18,21)(18,19)(21,20)(21,20)(21,20)
\psecurve(25,17)(25,17)(25,14)(27,15)(27,15)(27,15)
\psecurve(31,12)(31,12)(31,10)(34,11)(34,11)(34,11)
\psecurve(38,7)(38,7)(38,5)(40,6)(40,6)(40,6)
\psecurve(80,7)(80,7)(80,5)(78,6)(78,6)(78,6)
\psecurve(89,12)(89,12)(89,10)(86,11)(86,11)
\psecurve(95,17)(95,17)(95,14)(93,15)(93,15)(93,15)
\psecurve(101,21)(101,21)(101,19)(98,20)(98,20)
\end{pspicture}
\caption{Pictorial representation of the BC vertex integrant up to the normalization factor $\prod_{i< j} d_{I_{ij}}^{-1}$.
The graph is made of 10 disconnected unknots colored with representations $I_{ij}$. In each loop is inserted a $S^3$
element of the form $x_ix_j^{-1}$.}
\label{graphBC}
\end{center}
\end{figure}

As a result, one immediately obtains the pictorial representation of the BC vertex amplitude which
is given by the product of the normalization factor $\prod_{i< j} d_{I_{ij}}^{-1}$ and the graph in Figure 
(\ref{graphBC}). The graph consists in 10 disconnected loops colored by representations $I_{ij}$
which makes obvious that the vertex amplitude integrand is, up to a normalization, the product of 10
characters $\chi_{I_{ij}}(x_{ij})$. 

\medskip

{\bf Remark 2.} 
There is another equivalent expression for the vertex amplitude which was very useful to study the
classical behavior of the vertex amplitude found by Freidel and Louapre \cite{asymptotics}. 
This formula will not be used in this paper but it is still
interesting to mention it at least to ask the question whether a similar formula exists for the EPR model.
This formula is based on the simple fact that the character $\chi_I(x)$ depends only on the conjugacy 
class $\theta \in [0,\pi]$ of 
$x=\Lambda h(\theta) \Lambda^{-1}$: $\Lambda \in SU(2)/U(1)$ and $h(\theta)$ is in the Cartan torus of $SU(2)$.
This fact leads after some calculations to an expression of the vertex amplitude as an integral over the 
conjugacy classes:
\ba
V_{BC}(I_{ij},\omega_i^{BC}) \; = \; 
\int (\prod_{i\neq j}  d\theta_{ij} \, \frac{\sin(d_{I_{ij}}\theta_{ij})}{d_{I_{ij}}}) \, \widetilde{C}(\theta_{ij})\;.
\ea 
The notation $\widetilde{C}$ holds for the ``Fourier transform" of the distribution $C$;
it is a distribution as well given by:
\ba
\widetilde{C}(\theta_{ij}) \; \equiv \; \frac{2^{10}}{\pi^{10}} 
\int (\prod_{i <j} \sin\theta_{ij} \, d\Lambda_{ij}) \; 
C(\Lambda_{ij} h(\theta_{ij}) \Lambda_{ij}^{-1}) \; = \; \delta(G[\cos(\theta_{ij})]) 
\ea
where $G$ holds for the Gramm matrix. 
Such a relation  is in fact a particular example of a much more general duality relation \cite{duality}.

\subsubsection{The Engle-Pereira-Rovelli model}
The BC model has been considered as the most promising Spin-Foam model for a long time:
its definition is simple, it has a quite appealing physical interpretation and admits the good classical
limit \cite{asymptotics,num,EugenioLeonardo} in the sense that the associated vertex amplitude tends to the Regge action in the classical limit, apart from a term due to degenerate contributions, and it was also successful in reproducing the correct asymptotic behavior of the diagonal components of the graviton propagator \cite{propagator, numeric}.
Nevertheless, it has been recently realized that the model does not satisfy the required properties to 
reproduce at the semi-classical limit the non-diagonal components of the propagator \cite{difficulties}.
The reasons of this failure have been deeply investigated and a quest for a new model have been started.
Recent researches have led to the so-called EPR model which has been argued to be a serious candidate. 
This Section is devoted to recall the basis of this model in the Euclidean sector with 
no Immirzi-Barbero parameter $\gamma=0$.

As in the BC framework, Engle, Pereira and Rovelli have proposed a formula for the vertex amplitude $V_{EPR}$ only. 
To construct $V_{EPR}$, one starts by coloring the faces of the 4-simplex by simple representations and the
tetrahedra $i$ by specific intertwiners denoted $\omega_i^{EPR}$.
We propose to define $\omega_i^{EPR}$ throught its associated vector $v_i^{EPR}$ according to the formula (\ref{omega}). 
To do so, to each simple representation $(I_{ij},I_{ij})$, 
we associate the projector $\mathbb I_{2I_{ij}}:U_{I_{ij}I_{ij}} \rightarrow U_{2I_{ij}}$ from
the $SU(2)\times SU(2)$ representation's vector space $U_{I_{ij}I_{ij}}$ into the vector space
of the $SO(3)$ representation of spin $2I_{ij}$. In the standard bra-ket notation, the projector reads
$\mathbb I_{2I_{ij}}= \sum_m \vert m \, 2I_{ij}\ra \la m \, 2I_{ij}\vert $; it is clear that it can be trivially
identified to its dual $\mathbb I_{2I_{ij}}^*=\mathbb I_{2I_{ji}}$. 
Then, the vector $v_i$ is constructed from this projector as follows:
\ba\label{viEPR}
v_i^{EPR} \; \equiv \; \iota_\epsilon(\alpha_i) \, 
(\bigotimes_{j\neq i} \mathbb I_{2I_{ij}} ) 
\ea
where $\iota_\epsilon(\alpha_i)$ is a $SO(3)$ intertwiner, viewed as an element of the tensor product 
$\otimes_{j\neq i}V_{2I_{ij}}*$, caracterized by $\epsilon \in \{0,+,-\}$ and the
$SO(3)$ representation $\alpha_i$ as illustrated in the figure (\ref{intertwiners basis}).
As the vector $v_i^{EPR}$ is totally determined 
by a $SO(3)$ representation $\alpha_i$ and a choice of basis $\epsilon$,
we will identify  in the sequel the vector $v_i^{EPR}$ with the couple $(\alpha_i,\epsilon)$.
The pictorial representation of $v_i$ is the following: 
\be
\ifx\JPicScale\undefined\def\JPicScale{1}\fi
\psset{unit=\JPicScale mm}
\psset{linewidth=0.3,dotsep=1,hatchwidth=0.3,hatchsep=1.5,shadowsize=1,dimen=middle}
\psset{dotsize=0.7 2.5,dotscale=1 1,fillcolor=black}
\psset{arrowsize=1 2,arrowlength=1,arrowinset=0.25,tbarsize=0.7 5,bracketlength=0.15,rbracketlength=0.15}
\begin{pspicture}(0,0)(83,17)
\pspolygon[](2,10.4)(34,10.4)(34,5.6)(2,5.6)
\psline(2,12.8)(2,10.4)
\psline(5.2,12.8)(5.2,10.4)
\psline(11.6,12.8)(11.6,10.4)
\psline(14.8,12.8)(14.8,10.4)
\psline(21.2,12.8)(21.2,10.4)
\psline(24.4,12.8)(24.4,10.4)
\psline(30.8,12.8)(30.8,10.4)
\psline(34,12.8)(34,10.4)
\rput(18,8){{\small$v_i^{EPR}$}}
\rput(40,8){=}
\psecurve(45.6,14.6)(45.6,14.6)(47.2,9.8)(48.8,14.6)(48.8,14.6)(48.8,14.6)
\psecurve(56.6,14.6)(56.6,14.6)(58.2,9.8)(59.8,14.6)(59.8,14.6)(59.8,14.6)
\psecurve(67.6,14.6)(67.6,14.6)(69.2,9.8)(70.8,14.6)(70.8,14.6)(70.8,14.6)
\psecurve(78.4,14.8)(78.4,14.8)(80,10)(81.6,14.8)(81.6,14.8)(81.6,14.8)(81.6,14.8)
\psline(47.2,9.8)(53,5)
\psline(58.2,9.8)(53,5)
\psline(69.2,9.8)(74,5)
\psline(80.2,9.8)(74,5)
\rput(1,17){$I_{ij}$}
\rput(11,17){$I_{ik}$}
\rput(6,17){$I_{ij}$}
\rput(16,17){$I_{ik}$}
\rput(21,17){$I_{il}$}
\rput(25,17){$I_{il}$}
\rput(30,17){$I_{im}$}
\rput(35,17){$I_{im}$}
\rput(45,17){$I_{ij}$}
\rput(56,17){$I_{ik}$}
\rput(50,17){$I_{ij}$}
\rput(61,17){$I_{ik}$}
\rput(67,17){$I_{il}$}
\rput(72,17){$I_{il}$}
\rput(78,17){$I_{im}$}
\rput(83,17){$I_{im}$}
\rput(83,7){$2I_{im}$}
\rput(68,7){$2I_{il}$}
\rput(45,7){$2I_{ij}$}
\rput(60,7){$2I_{ik}$}
\rput(63,0){$\alpha_{\st i}$}
\psecurve(74,5)(74,5)(63,2)(53,5)(53,5)(53,5)
\end{pspicture}
\ee
Note that we made a particular choice for $\epsilon$ to draw the picture; 
another choice would lead to a different contraction of the four edges colored by the representations $2I_{ij}$.
Contrary to the BC model, the EPR intertwiner between four given representations $I_{ij}$ is not unique for it depends
on $\alpha_i$ and $\varepsilon$, both belonging to a finite set.

\begin{figure}
\begin{center}
\ifx\JPicScale\undefined\def\JPicScale{0.8}\fi
\psset{unit=\JPicScale mm}
\psset{linewidth=0.3,dotsep=1,hatchwidth=0.3,hatchsep=1.5,shadowsize=1,dimen=middle}
\psset{dotsize=0.7 2.5,dotscale=1 1,fillcolor=black}
\psset{arrowsize=1 2,arrowlength=1,arrowinset=0.25,tbarsize=0.7 5,bracketlength=0.15,rbracketlength=0.15}
\begin{pspicture}(0,0)(40.25,73.88)
\psline[linewidth=0.2](24,6)(11.5,17.25)
\psline[linewidth=0.2](24,6)(36.5,17.25)
\psline[linewidth=0.2](24,6)(24,73.5)
\psline[linewidth=0.2](11.5,62.25)(24,73.5)
\psline[linewidth=0.2](36.5,62.25)(24,73.5)
\psline[linewidth=0.2](11.5,33)(11.5,48)
\psline[linewidth=0.2](11.5,33)(36.5,17.25)
\psline[linewidth=0.2](36.5,49.5)(36.5,33)
\psline[linewidth=0.2,border=0.3](11.5,17.25)(36.5,33)
\psline[linewidth=0.2,border=0.3](11.5,62.25)(36.5,48.75)
\psline[linewidth=0.2,border=0.3](11.5,48.75)(36.5,62.25)
\psline[linewidth=0.2](11.5,61.5)(11.5,48)
\psline[linewidth=0.2](36.47,62.33)(36.5,49.5)
\psline[linewidth=0.2](11.5,33)(11.5,18)
\psline[linewidth=0.2](36.5,33)(36.47,17.2)
\rput(27.13,40.42){$i$}
\rput(40.25,40.5){$i_+$}
\rput(15.25,40.5){$i_-$}
\rput(14,70){$2j_1$}
\rput(34,70){$2j_2$}
\rput(14,9){$2j_3$}
\rput(34,9){$2j_4$}
\rput(9,54){$j_1$}
\rput(18,49){$j_2$}
\rput(39,54){$j_2$}
\rput(29,49){$j_1$}
\psline[linewidth=0.2](11.5,31.5)(11.47,17.33)
\psline[linewidth=0.2](36.5,33)(36.5,19.5)
\rput(9,24){$j_3$}
\rput(18,32){$j_4$}
\rput(39,24){$j_4$}
\rput(30,32){$j_3$}
\rput{90}(11.65,62.32){\psellipse[fillstyle=solid](0,0)(0.46,0.43)}
\rput{90}(24.03,73.42){\psellipse[fillstyle=solid](0,0)(0.46,0.43)}
\rput{90}(36.41,62.36){\psellipse[fillstyle=solid](0,0)(0.46,0.43)}
\rput{90}(36.56,48.92){\psellipse[fillstyle=solid](0,0)(0.46,0.43)}
\rput{90}(11.57,48.71){\psellipse[fillstyle=solid](0,0)(0.46,0.43)}
\rput{90}(11.59,32.91){\psellipse[fillstyle=solid](0,0)(0.46,0.43)}
\rput{90}(11.6,17.11){\psellipse[fillstyle=solid](0,0)(0.46,0.43)}
\rput{90}(23.98,5.91){\psellipse[fillstyle=solid](0,0)(0.46,0.43)}
\rput{90}(36.43,17.08){\psellipse[fillstyle=solid](0,0)(0.46,0.43)}
\rput{90}(36.51,32.92){\psellipse[fillstyle=solid](0,0)(0.46,0.43)}
\end{pspicture}
\caption{EPR fusion coefficients. The edges are colored with $SU(2)$ representations and the vertices with 
symmetric $SU(2)$ 3j symbols. The picture illustrates the coefficient $f(\omega_i,I_{ij},\iota_\epsilon(\alpha))$
for $I_{ij}=\{j_1,j_2,j_3,j_4\}$, $\omega_i$ is caracterized by $i$ (and some $\epsilon$) and $\alpha=(i_+,i_-)$.}
\label{functionf}
\end{center}
\end{figure}
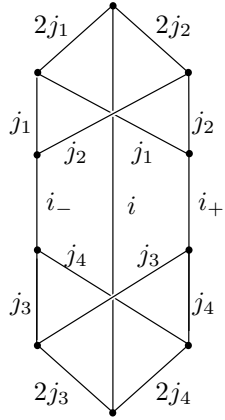

Now, it is possible to decompose the EPR intertwiner in any tensor product basis of the space of 4-valent 
$SU(2)\times SU(2)$ intertwiners. We are interested in its decomposition in the basis of the type $(\epsilon,\epsilon)$
whose elements are denoted $\iota_\epsilon(\alpha)$
After some simple calculation, we recover the following expression of the EPR intertwiner given in the literature:
\ba\label{EPRvertex1}
\omega_i^{EPR} \; = \; \sum_{\alpha} 
f(\omega_i, I_{ij}, \iota_\epsilon(\alpha)) \, \iota_\epsilon(\alpha)  
\ea
where the coefficient $f$ is graphically ``represented" in the Figure (\ref{functionf})
 and
the sum is finite and runs over $SU(2)\times SU(2)$ representations $\alpha$ with a fixed chosen 
basis $\epsilon$. In the notation of Engle-Pereira-Rovelli, $\alpha$ is denoted $(i_+,i_-)$ and
the representation defining $\omega_i$ is denoted $i$. 
Note that the sum (\ref{EPRvertex1}) is not restricted to simple representations.

\medskip

Now, we have all the ingredients to compute the vertex amplitude $V_{EPR}(I_{ij},\omega_i^{EPR})$ for the $EPR$ model.
>From the expression (\ref{EPRvertex1}), we show immediately that:
\ba
V_{EPR}(I_{ij},\omega_i^{EPR}) \; = \; \sum_{\alpha=(i_+,i_-)} 
f(\omega_i, I_{ij}, \iota_\epsilon(i_+,i_-))   \; V_{BF}(I_{ij},i_+)\; V_{BF}(I_{ij},i_-)
\ea
where $V_{BF}(I_{ij},i_\pm)$ are the $SU(2)$ 15j symbols which depends on the representations
$I_{ij}$ and $\alpha$ but also on the choice of the basis $\epsilon$ which has not been explicitely written.
The sums runs over $SU(2)\times SU(2)$ representations $\alpha$ with a fixed $\epsilon$. 
Such a formula is rather complicated and one might prefer working instead with an integral formula
of the form (\ref{general formula}). 
To obtain such a formula, one has to separate in the integral (\ref{general formula integral}) 
the variables $u_i$ from the variables
$x_i$ as in (\ref{Vema}) and then to perform the integration over the variables $u_i$.
These last integrations are very simple to compute: the integration over $u_3$ is trivial
and those over the remaining  variables $u_i$ give a simple normalisation factor 
$N=(d_{2I_{12}}d_{2I_{45}}d_{\omega_1}d_{\omega_5})^{-1}$. 

\begin{figure}
\begin{center}
\ifx\JPicScale\undefined\def\JPicScale{0.6}\fi
\psset{unit=\JPicScale mm}
\psset{linewidth=0.3,dotsep=1,hatchwidth=0.3,hatchsep=1.5,shadowsize=1,dimen=middle}
\psset{dotsize=0.7 2.5,dotscale=1 1,fillcolor=black}
\psset{arrowsize=1 2,arrowlength=1,arrowinset=0.25,tbarsize=0.7 5,bracketlength=0.15,rbracketlength=0.15}
\begin{pspicture}(0,0)(120.06,118)
\psecurve(45,109)(45,109)(35.3,87.67)(7,81)(7,81)(7,81)
\psecurve(48,109)(48,109)(37.9,86.33)(7,78)(7,78)(7,78)
\psecurve(55,109)(55,109)(49.6,65)(25,17)(25,17)
\psecurve(57,109)(57,109)(52.2,63.67)(27,15)(27,15)
\psecurve(63,109)(63,109)(67.8,63.67)(93,15)(93,15)(93,15)
\psecurve(65,109)(65,109)(70.4,65)(95,17)(95,17)(95,17)
\psecurve(72,109)(72,109)(82.1,86.33)(113,78)(113,78)(113,78)
\psecurve(75,109)(75,109)(84.7,87.67)(113,81)(113,81)(113,81)
\psline[border=0.3](7,73)(113,73)
\psline[border=0.3](7,70)(113,70)
\psecurve(18,21)(18,21)(24.9,42.33)(7,52)(7,52)(7,52)
\psecurve(21,20)(21,20)(27.5,43.67)(7,55)(7,55)(7,55)
\psecurve[border=0.3,curvature=1.0 0.1 0.0](7,64)(7,64)(53.5,51.67)(89,12)(89,12)(89,12)
\psecurve[border=0.3,curvature=1.0 0.1 0.0](7,61)(7,61)(53.5,47.67)(86,11)(86,11)(86,11)
\psecurve[border=0.3,curvature=1.0 0.1 0.0](113,64)(113,64)(66.5,51.67)(31,12)(31,12)(31,12)
\psecurve[border=0.3,curvature=1.0 0.1 0.0](113,61)(113,61)(66.5,47.67)(34,11)(34,11)(34,11)
\psecurve(113,55)(113,55)(92.5,43.67)(98,20)(98,20)(98,20)
\psecurve(113,52)(113,52)(95.1,42.33)(101,21)(101,21)(101,21)
\psecurve(38,7)(38,7)(60,26.33)(80,7)(80,7)(80,7)
\psecurve(40,6)(40,6)(60,23.67)(78,6)(78,6)(78,6)
\rput{90}(39.85,88.33){\psellipse[fillcolor=darkgray,fillstyle=solid](0,0)(2,-1.95)}
\rput{90}(51.55,61.67){\psellipse[fillcolor=darkgray,fillstyle=solid](0,0)(2,-1.95)}
\rput{90}(82.1,91){\psellipse[fillcolor=darkgray,fillstyle=solid](0,0)(2,-1.95)}
\rput{90}(71.7,62.33){\psellipse[fillcolor=darkgray,fillstyle=solid](0,0)(2,-1.95)}
\rput(44.4,83.67){{\tiny$x_1x_2^{-1}$}}
\rput{90}(51.55,48.33){\psellipse[fillcolor=darkgray,fillstyle=solid](0,0)(2,-1.95)}
\rput{90}(67.8,47.67){\psellipse[fillcolor=darkgray,fillstyle=solid](0,0)(2,-1.95)}
\rput{90}(24.25,41){\psellipse[fillcolor=darkgray,fillstyle=solid](0,0)(2,-1.95)}
\rput{90}(60,23){\psellipse[fillcolor=darkgray,fillstyle=solid](0,0)(2,-1.95)}
\rput{90}(96.4,43.67){\psellipse[fillcolor=darkgray,fillstyle=solid](0,0)(2,-1.95)}
\rput{90}(60,69.67){\psellipse[fillcolor=darkgray,fillstyle=solid](0,0)(2,-1.95)}
\rput(57,57){{\tiny $x_1 x_3^{-1}$}}
\rput(80,62){{\tiny $x_1 x_4^{-1}$}}
\rput(91,91.67){{\tiny $x_1 x_5^{-1}$}}
\rput(60,65){{\tiny $x_2 x_5^{-1}$}}
\rput(50.0,43.67){{\tiny $x_2 x_4^{-1}$}}
\rput(16.4,39.67){{\tiny $x_2 x_3^{-1}$}}
\rput(60,18.33){{\tiny $x_3 x_4^{-1}$}}
\rput(70.4,43.67){{\tiny $x_3 x_5^{-1}$}}
\rput(101.6,39.67){{\tiny $x_4 x_5^{-1}$}}
\psecurve(45,109)(45,109)(47,111)(48,109)(48,109)(48,109)
\psecurve(55,109)(55,109)(56,111)(57,109)(57,109)(57,109)(57,109)(57,109)
\psecurve(63,109)(63,109)(64,111)(65,109)(65,109)(65,109)
\psecurve(72,109)(72,109)(73,111)(75,109)(75,109)(75,109)
\psecurve(113,81)(113,81)(115,79)(113,78)(113,78)(113,78)
\psecurve(113,73)(113,73)(115,71)(113,70)(113,70)(113,70)
\psecurve(113,64)(113,64)(115,62)(113,61)(113,61)(113,61)
\psecurve(113,55)(113,55)(115,53)(113,52)(113,52)
\psecurve(7,52)(7,52)(5,54)(7,55)(7,55)(7,55)
\psecurve(7,61)(7,61)(5,63)(7,64)(7,64)(7,64)
\psecurve(7,70)(7,70)(5,72)(7,73)(7,73)(7,73)
\psecurve(7,78)(7,78)(5,80)(7,81)(7,81)(7,81)
\psecurve(18,21)(18,21)(18,19)(21,20)(21,20)(21,20)
\psecurve(25,17)(25,17)(25,14)(27,15)(27,15)(27,15)
\psecurve(31,12)(31,12)(31,10)(34,11)(34,11)(34,11)
\psecurve(38,7)(38,7)(38,5)(40,6)(40,6)(40,6)
\psecurve(80,7)(80,7)(80,5)(78,6)(78,6)(78,6)
\psecurve(89,12)(89,12)(89,10)(86,11)(86,11)
\psecurve(95,17)(95,17)(95,14)(93,15)(93,15)(93,15)
\psecurve(101,21)(101,21)(101,19)(98,20)(98,20)
\psecurve(47,111)(47,111)(52,116)(56,111)(56,111)(56,111)(56,111)
\psecurve(64,111)(64,111)(68,116)(73,111)(73,111)(73,111)
\psecurve(52,116)(52,116)(60,118)(68,116)(68,116)(68,116)
\psecurve(115,79)(115,79)(117,75)(115,71)(115,71)(115,71)
\psecurve(115,62)(115,62)(117,58)(115,54)(115,54)(115,54)
\psecurve(117,75)(117,75)(120,66)(117,58)(117,58)(117,58)
\psecurve(5,54)(5,54)(3,59)(5,63)(5,63)(5,63)
\psecurve(5,72)(5,72)(3,76)(5,80)(5,80)(5,80)
\psecurve(3,59)(3,59)(0,68)(3,76)(3,76)(3,76)
\psecurve(100.98,19.11)(100.98,19.11)(99.35,14.95)(95.03,13.76)(95.03,13.76)(95.03,13.76)
\psecurve(89,10)(89,10)(85,5)(80,5)(80,5)(80,5)
\psecurve(99,15)(99,15)(94.65,8.05)(84.65,5.05)(84.65,5.05)(84.65,5.05)
\psecurve(38,5)(38,5)(33.42,5.61)(31,10)(31,10)(31,10)
\psecurve(25,14)(25,14)(19.68,14.85)(18,19)(18,19)(18,19)
\psecurve(34,5)(34,5)(25,8)(19.68,14.85)(19.68,14.85)(19.68,14.85)
\end{pspicture}
\caption{Picturial representation of the EPR argument in the integral formula: vertices
are labelled by $i=1,\cdots,5$ where $i=1$ is the top vertex and the others are enumerated according to
the anti-clockwise orientation; edges are then oriented and are labelled by $(ij)$ with $i<j$.
The doubled lines are colored
with simple representations $(I_{ij},I_{ij})$. The lines $(ij)$ in the same pair are linked to a line colored with the 
representation $2I_{ij}$. At each vertex, the four single lines are linked with a line of representation $\omega_i$.}
\label{Epr integrand}
\end{center}
\end{figure}
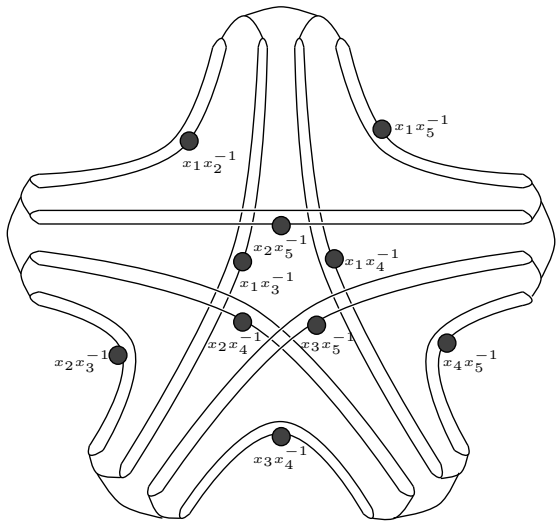

Afterwards, the vertex amplitude reduces to the formula:
\ba\label{simplifiedintegralEPR}
V_{EPR}(I_{ij},\omega_i) \; = \; N \,
\int \prod_{i\neq j} dx_{ij} \, C(x_{ij}) \, {\mathcal V}(I_{ij},\omega_i;x_{ij})
\ea
where the amplitude ${\cal V}$ is a function of the 10 variables $x_{ij}$ and is graphically represented in the
Figure (\ref{Epr integrand}).
This formula is the EPR counterpart of the formula (\ref{VBCintegral}) for the BC model. 
It will appear very useful in the next Section to make a contact with Loop Quantum Gravity.
It might also be useful to study the classical and semi-classical properties of the EPR model
as it is the case for the BC model.

\subsubsection{A direct generalization: the Freidel-Krasnov models}
This Section is devoted to present a very direct generalization of the EPR model. 
This generalization leads to a large class of Spin-Foam models to which belong both the EPR and the 
BC models. 

To motivate the construction of FK models, 
let us recall that the vector $v_i^{EPR}$, necessary to define the EPR intertwiner $\omega_i^{EPR}$,
has been constructed making use of a projector $\mathbb I_{2I_{ij}}$ from the vector space of the $SU(2)\times SU(2)$
simple representation $(I_{ij},I_{ij})$ into the SO(3) vector space representation $U_{2I_{ij}}$. 
A direct generalization would be to define
a vector $v_i^{gen}$ using instead, at each vertex $i$,  
projectors $\mathbb I_{K^i_j}$ from $V_{I_{ij}I_{ij}}$ into the SO(3) 
representation $U_{K^i_j}$ for any representation $K^i_j \in [0,2I_{ij}]$. The formal expression of the general
vector is then the following:
\ba
v_i^{gen} \; \equiv \; \iota_\epsilon(\alpha_i) \, 
(\bigotimes_{j\neq i} \mathbb K^i_j)  \;.
\ea
The vector $v_i^{gen}$ so defined depends on the choice of the intertwiner $\iota_\epsilon(\alpha_i)$
and on the representations $K^i_j$. It is represented by the following diagram
\be
\ifx\JPicScale\undefined\def\JPicScale{1}\fi
\psset{unit=\JPicScale mm}
\psset{linewidth=0.3,dotsep=1,hatchwidth=0.3,hatchsep=1.5,shadowsize=1,dimen=middle}
\psset{dotsize=0.7 2.5,dotscale=1 1,fillcolor=black}
\psset{arrowsize=1 2,arrowlength=1,arrowinset=0.25,tbarsize=0.7 5,bracketlength=0.15,rbracketlength=0.15}
\begin{pspicture}(0,0)(83,17)
\pspolygon[](2,10.4)(34,10.4)(34,5.6)(2,5.6)
\psline(2,12.8)(2,10.4)
\psline(5.2,12.8)(5.2,10.4)
\psline(11.6,12.8)(11.6,10.4)
\psline(14.8,12.8)(14.8,10.4)
\psline(21.2,12.8)(21.2,10.4)
\psline(24.4,12.8)(24.4,10.4)
\psline(30.8,12.8)(30.8,10.4)
\psline(34,12.8)(34,10.4)
\rput(18,8){{\small$v_i^{gen}$}}
\rput(40,8){=}
\psecurve(45.6,14.6)(45.6,14.6)(47.2,9.8)(48.8,14.6)(48.8,14.6)(48.8,14.6)
\psecurve(56.6,14.6)(56.6,14.6)(58.2,9.8)(59.8,14.6)(59.8,14.6)(59.8,14.6)
\psecurve(67.6,14.6)(67.6,14.6)(69.2,9.8)(70.8,14.6)(70.8,14.6)(70.8,14.6)
\psecurve(78.4,14.8)(78.4,14.8)(80,10)(81.6,14.8)(81.6,14.8)(81.6,14.8)(81.6,14.8)
\psline(47.2,9.8)(53,5)
\psline(58.2,9.8)(53,5)
\psline(69.2,9.8)(74,5)
\psline(80.2,9.8)(74,5)
\rput(1,17){$J_{ij}$}
\rput(11,17){$J_{ik}$}
\rput(6,17){$J_{ij}$}
\rput(16,17){$J_{ik}$}
\rput(21,17){$J_{il}$}
\rput(25,17){$J_{il}$}
\rput(30,17){$J_{im}$}
\rput(35,17){$J_{im}$}
\rput(45,17){$J_{ij}$}
\rput(56,17){$J_{ik}$}
\rput(50,17){$J_{ij}$}
\rput(61,17){$J_{ik}$}
\rput(67,17){$J_{il}$}
\rput(72,17){$J_{il}$}
\rput(78,17){$J_{im}$}
\rput(83,17){$J_{im}$}
\rput(83,7){$K^i_{m}$}
\rput(68,7){$K^i_{l}$}
\rput(45,7){$K^i_{j}$}
\rput(60,7){$K^i_{k}$}
\rput(63,0){$\alpha_i$}
\psecurve(74,5)(74,5)(63,2)(53,5)(53,5)(53,5)
\end{pspicture}
\ee
This leads to a vertex amplitude very similar to the EPR one. In particular, its integral formula takes the same
form of (\ref{simplifiedintegralEPR}) where the normalization factor is changed into 
$N=(d_{I^1_2} d_{I^4_5}d_{\omega_1}d_{\omega_1})^{-1}$ and the function ${\cal V}$ is represented
by the same graph drawn in the Figure (\ref{Epr integrand}) with different spin labels.

As a consequence, we get a large class of Spin-Foam models vertex amplitudes
$V_i^{gen}$ which depends not only on the 10 representations $I_{ij}$ coloring the faces of the 4-simplex
but also depends on 5 other representations per tetrahedron $i$ which have been denoted $\alpha_i$,
$K^i_j$. Up to now, only special cases of such models have been studied:
the BC model where $K^i_j=\alpha_i=0$, the EPR model where $K^i_j=K^j_i=2I_{ij}$ and $\alpha_i$
is a free parameter. Thus, either we choose to project into the trivial representation either into
the hightest representation. The FK model consists in another choice of the representations $K^i_j$
and $\alpha_i$. 

Many arguments lead to the fact that the EPR intertwiners define the good physical model, namely the one which
should reproduce the discretized path integral of the Euclidean Plebanski theory.

\section{The vertex and the physical scalar product}
In this Section we are proposing a link between (covariant) Spin-Foam models and (canonical) Loop Quantum Gravity.
To explain our strategy,
we start by recalling some needed basic results of LQG. 
One of the main points of LQG is the assumption that physical states can be constructed from
the so-called kinematical Hilbert space ${\cal H}_{kin}$ which consists in the space
of cylindrical functions endowed with the kinematical scalar product $\langle,\rangle$ defined from the 
$SU(2)$ Haar measure. The Spin-network states form an orthonormal  basis of ${\cal H}_{kin}$.
Then, the idea is basicly to impose the constraints of gravity to extract physical states out of the kinematical
space. So far, we know how to impose the Gauss constraint and the space-diffeomorphisms constraints and this leads
to the construction of the the diffeomorphism invariant states: they form 
the space ${\cal H}_{diff}$ which is endowed with the Ashtekar-Lewandowski measure \cite{AL}.
The physical Hilbert space ${\cal H}_{phys}$ is still unknown but expected to be constructed from 
the Ashtekar-Lewandowski measure. Up to now, we do not how to solve the remaining Hamiltonian constraint.
Spin-Foam models have been introduced as an alternative to find physical states and
the physical scalar product in the sense that the amplitude of a Spin-Foam models should reproduce
the physical scalar product between the states at the boundary of the Spin-Foam. This Section aims 
precisely at clarifying this last point in a simple case.

\begin{figure}[h]
\begin{center}
\ifx\JPicScale\undefined\def\JPicScale{1}\fi
\psset{unit=\JPicScale mm}
\psset{linewidth=0.3,dotsep=1,hatchwidth=0.3,hatchsep=1.5,shadowsize=1,dimen=middle}
\psset{dotsize=0.7 2.5,dotscale=1 1,fillcolor=black}
\psset{arrowsize=1 2,arrowlength=1,arrowinset=0.25,tbarsize=0.7 5,bracketlength=0.15,rbracketlength=0.15}
\begin{pspicture}(0,0)(51,50)
\psline[linewidth=0.2,border=0.3](36,33)(44,3)
\psline[linewidth=0.2,border=0.3](51,28)(17,3)
\psline[linewidth=0.2,border=0.3](43.67,3)(8,28)
\psline[linewidth=0.2,border=0.3](8,28)(51,28)
\psline[linewidth=0.2,border=0.3](17,3)(24,33)
\psline[linewidth=0.2](51,28)(40,41)
\psline[linewidth=0.2](48,49)(40,42)
\psline[linewidth=0.2,border=0.3](36,34)(50,41)
\psline[linewidth=0.2](20,41)(8,28)
\psline[linewidth=0.2,border=0.3](10,40)(24,34)
\psline[linewidth=0.2](12,49)(20,42)
\psline[linewidth=0.2](8,28)(17,3)
\psline[linewidth=0.2](17.33,3)(44,3)
\psline[linewidth=0.2](44,3)(51,28)
\rput{90}(38.05,24.06){\psellipse[fillstyle=solid](0,0)(1.06,1.05)}
\rput{90}(21.95,24.06){\psellipse[fillstyle=solid](0,0)(1.06,1.05)}
\rput{90}(25.95,15.06){\psellipse[fillstyle=solid](0,0)(1.06,1.05)}
\rput{90}(33.95,15.06){\psellipse[fillstyle=solid](0,0)(1.06,1.05)}
\rput{90}(47.95,17.06){\psellipse[fillstyle=solid](0,0)(1.06,1.05)}
\rput{90}(12,16.94){\psellipse[fillstyle=solid](0,0)(1.06,1.05)}
\rput{90}(30.95,3.06){\psellipse[fillstyle=solid](0,0)(1.06,1.05)}
\rput{90}(30.05,28.06){\psellipse[fillstyle=solid](0,0)(1.06,1.05)}
\rput(9,37){\tiny$z_{2}$}
\rput(11,46){\tiny$z_{1}$}
\rput(50,37){\tiny$z_{3}$}
\rput(49,46){\tiny$z_{4}$}
\rput{90}(11.95,32.94){\psellipse[fillstyle=solid](0,0)(1.05,1.05)}
\rput{90}(46.95,32.94){\psellipse[fillstyle=solid](0,0)(1.06,1.05)}
\rput(8,32){\tiny$x_{21}$}
\rput(25,22){\tiny$x_{31}$}
\rput(35,22){\tiny$x_{41}$}
\rput(51,32){\tiny$x_{51}$}
\rput(23,13){\tiny$x_{24}$}
\rput(37,13){\tiny$x_{35}$}
\rput(31,0){\tiny$x_{34}$}
\rput(10,14){\tiny$x_{23}$}
\rput(50,14){\tiny$x_{45}$}
\rput(30,31){\tiny$x_{25}$}
\psline[linewidth=0.2,fillstyle=solid](24,34)(31,49)
\psline[linewidth=0.2,fillstyle=solid](31,49)(36,34)
\psline[linewidth=0.2,fillstyle=solid](31,49)(40,42)
\psline[linewidth=0.2,fillstyle=solid](31,49)(20,42)
\rput{90}(34.95,45.94){\psellipse[fillstyle=solid](0,0)(1.06,1.05)}
\rput(38,50){\tiny$y_{5}$}
\rput{90}(26.95,45.94){\psellipse[fillstyle=solid](0,0)(1.06,1.05)}
\rput(23,47){\tiny$y_{2}$}
\rput{90}(26.95,38.94){\psellipse[fillstyle=solid](0,0)(1.06,1.05)}
\rput{90}(34.05,39.06){\psellipse[fillstyle=solid](0,0)(1.06,1.05)}
\rput(23,40){\tiny$y_{3}$}
\rput(38,40){\tiny$y_{4}$}
\psline[linewidth=0.2](12,48)(20,41)
\rput{90}(14.05,46.94){\psellipse[fillcolor=lightgray,fillstyle=solid](0,0)(1.06,1.05)}
\psline[linewidth=0.2,border=0.3](24,33)(10,39)
\rput{90}(11.95,38.94){\psellipse[fillcolor=lightgray,fillstyle=solid](0,0)(1.05,1.05)}
\psline[linewidth=0.2,border=0.3](36,33)(50,40)
\rput{90}(47.95,39.94){\psellipse[fillcolor=lightgray,fillstyle=solid](0,0)(1.06,1.05)}
\psline[linewidth=0.2](48,48)(40,41)
\rput{90}(46.05,46.94){\psellipse[fillcolor=lightgray,fillstyle=solid](0,0)(1.06,1.05)}
\end{pspicture}
\caption{Representation of the graph $\widetilde{\Gamma}$. The subgraphs associated to $\tau_1$
and $\tau_4$ have been underlined and the group variables associated to each edge have been emphasized.}
\label{Gammaema}
\end{center}
\end{figure}
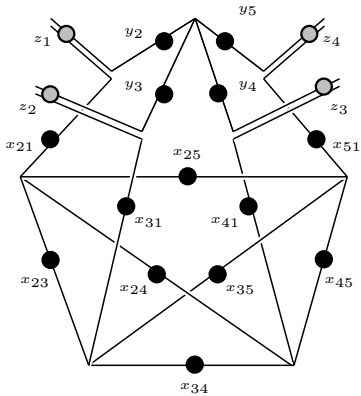

More precisely, we consider the Spin-Foam associated to the 4-simplex graph denoted $\Gamma$. Its amplitude
is given, up to some eventual irrelevant normalization factors, by the vertex amplitude $V$. From the general
boundary (covariant) formulation point of view, $\Gamma$ is viewed as a graph interpolating 
between two kinematical boundary states
which are $\tau_1$ and $\tau_4$ as schematically depicted in the figure (\ref{Gammaema}). 
In fact, as shown in the figure (\ref{Gammaema}), $\tau_1$ and $\tau_4$ belong to the space 
$\text{Cyl}(\widetilde{\Gamma})$ where $\widetilde{\Gamma}$ is the union of $\Gamma$ with four free ends. 
These free ends have been added for technical purposes only. 
Notice that $\Gamma$
can be equivalently interpreted as the graph interpolating between two different graphs that would be
denoted $\tau_2$ (with two vertices) and $\tau_3$ (with three vertices). For that, one would need to introduce also
some free ends at the graph $\Gamma$.
From the canonical point of view, the states $\tau_1$ and $\tau_4$ are considered schematically
as cylindrical functions on the graph $\widetilde{\Gamma}$. Therefore, one naturally asks the question whether it exists a  
``physical projector" P acting on the space $\text{Cyl}(\widetilde{\Gamma})$ such that its matrix element 
$\la \tau_4,P\tau_1\ra$ constructed from the kinematical scalar product gives the vertex amplitude.
The notation $\la \tau_4,P\tau_1\ra$ can be misleading because $P$ has in fact to be viewed as
a state in the sense of Gelfand-Naimark-Segal (GNS), i.e. $P$ is a linear form on $\text{Cyl}(\widetilde{\Gamma})$
and the physical scalar product reads $\la \tau_4,P\tau_1\ra=P(\overline{\tau_4}\tau_1)$. 
We abusively use
the same notation for the projector viewed as a `` matricial operator" or a linear form.  
To be interpreted as a GNS state, $P$ has to satisfy additional properties, like the positivity,
that we will not discuss here.
We show that it is possible to construct explicitely such an operator $P$ for the topological, the BC and 
the EPR models. The ``projector" for the FK model can also be obtained immediately generalizing the
construction in the EPR case. We will use the obvious notations $P_{BF}$, $P_{BC}$ and $P_{EPR}$
to denote the physical ``projector" in the different cases.

There are two important points to clarify. The first one is the issue of uniqueness of the solution:
we find one (class of) solution(s) for $P$ in each model but we do not know if it is unique 
(in some precise sense of course). 
Second we work in the kinematical Hilbert space and we expect $P$ to behave correctly with respect to
diffeomorphisms invariance in order to extend it to ${\cal H}_{diff}$. 
We hope to address these important mathematical issues in the future.

\subsection{The topological model}
The topological model is the simplest case to consider. Even if it is not of a great physical interest, it is
a good toy model to test the possibility of constructing a ``physical projector" $P$. Furthermore, we will see 
that this construction will be useful to study the other more physical cases. 
Let us emphasize that the construction of $P_{BF}$ is very similar to the construction of 
the projector into physical states in three dimensions as expected from the topological nature of the model.

\begin{figure}[h]
\begin{center}
\ifx\JPicScale\undefined\def\JPicScale{1}\fi
\psset{unit=\JPicScale mm}
\psset{linewidth=0.3,dotsep=1,hatchwidth=0.3,hatchsep=1.5,shadowsize=1,dimen=middle}
\psset{dotsize=0.7 2.5,dotscale=1 1,fillcolor=black}
\psset{arrowsize=1 2,arrowlength=1,arrowinset=0.25,tbarsize=0.7 5,bracketlength=0.15,rbracketlength=0.15}
\begin{pspicture}(0,0)(51,49)
\psline[linewidth=0.2,border=0.3](36,34)(44,3)
\psline[linewidth=0.2,border=0.3](51,28)(17,3)
\psline[linewidth=0.2,border=0.3](43.67,3)(8,28)
\psline[linewidth=0.2,border=0.3](8,28)(51,28)
\psline[linewidth=0.2,border=0.3](17,3)(24,34)
\psline[linewidth=0.2](51,28)(40,42)
\psline(48,49)(40,42)
\psline[border=0.3](36,34)(50,41)
\psline[linewidth=0.2](20,42)(8,28)
\psline[border=0.3](10,40)(24,34)
\psline(12,49)(20,42)
\psline[linewidth=0.2](8,28)(17,3)
\psline[linewidth=0.2](17.33,3)(44,3)
\psline[linewidth=0.2](44,3)(51,28)
\rput{90}(38.05,24.06){\psellipse[fillstyle=solid](0,0)(1.05,1.05)}
\rput{90}(45.95,47.06){\psellipse[fillcolor=lightgray,fillstyle=solid](0,0)(1.05,1.05)}
\rput{90}(13.95,47){\psellipse[fillcolor=lightgray,fillstyle=solid](0,0)(1.05,1.05)}
\rput{90}(11.95,38.94){\psellipse[fillcolor=lightgray,fillstyle=solid](0,0)(1.05,1.05)}
\rput{90}(47.95,39.94){\psellipse[fillcolor=lightgray,fillstyle=solid](0,0)(1.05,1.05)}
\rput{90}(21.95,24.06){\psellipse[fillstyle=solid](0,0)(1.05,1.05)}
\rput{90}(25.95,15.06){\psellipse[fillstyle=solid](0,0)(1.05,1.05)}
\rput{90}(33.95,15.06){\psellipse[fillstyle=solid](0,0)(1.05,1.05)}
\rput{90}(47.95,17.06){\psellipse[fillstyle=solid](0,0)(1.05,1.05)}
\rput{90}(12,16.94){\psellipse[fillstyle=solid](0,0)(1.05,1.05)}
\rput{90}(30.95,3.05){\psellipse[fillstyle=solid](0,0)(1.05,1.05)}
\rput{90}(30.05,28.06){\psellipse[fillstyle=solid](0,0)(1.05,1.05)}
\rput(9,37){\tiny$z_{2}$}
\rput(11,46){\tiny$z_{1}$}
\rput(50,37){\tiny$z_{3}$}
\rput(49,46){\tiny$z_{4}$}
\rput{90}(11.95,32.94){\psellipse[fillstyle=solid](0,0)(1.05,1.05)}
\rput{90}(46.95,32.94){\psellipse[fillstyle=solid](0,0)(1.05,1.05)}
\rput(8,32){\tiny$x_{21}$}
\rput(25,22){\tiny$x_{31}$}
\rput(35,22){\tiny$x_{41}$}
\rput(51,32){\tiny$x_{51}$}
\rput(23,13){\tiny$x_{24}$}
\rput(37,13){\tiny$x_{35}$}
\rput(31,0){\tiny$x_{34}$}
\rput(10,14){\tiny$x_{23}$}
\rput(50,14){\tiny$x_{45}$}
\rput(30,31){\tiny$x_{25}$}
\end{pspicture}
\caption{Pictorial representation of the graph associated to $\tau_1$ and $\tau_4$ separately.
The free edges are oriented from the vertices to the free ends; the internal edges are oriented according to the
order on the vertices.
The variables associated to the free ends are denoted $z_k$ for the two graphs; 
those associated to the internal edges of $\tau_4$ are  denoted $x_{ij}$ with $i,j=2,\cdots,5$;
those associated to the internal edges of $\tau_1$ are denoted $y_h$.}
\label{tau4}
\end{center}
\end{figure}
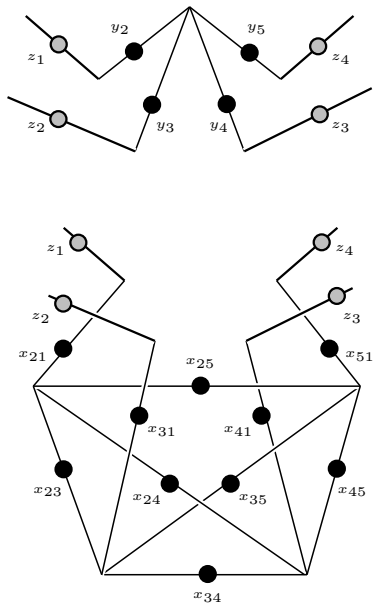

As we said in the introduction of this Section, the boundary states $\tau_1$ and $\tau_4$  
are elements of $\text{Cyl}(\widetilde{\Gamma})$: $\tau_1$ 
is a function of the eight group variables $y_h,z_k$, with $k =1,\cdots, 4$ and $h=2,\cdots,5$, 
as shown in the figure (\ref{tau4}); 
$\tau_4$ is a function of fourteen group variables, ten of them are denoted $x_{ij}$ with $i,j=1,\cdots,5$ and $i\neq j$, and the four remaining are the 
$z_k$ variables as shown in the figure \ref{tau4}. 
Note that the $z_k$ group variables are those associated to the free ends of $\tilde{\Gamma}$
which are common to the spin-network graphs associated to $\tau_1$ and $\tau_5$.

\begin{figure}[h]
\begin{center}
\ifx\JPicScale\undefined\def\JPicScale{1}\fi
\psset{unit=\JPicScale mm}
\psset{linewidth=0.3,dotsep=1,hatchwidth=0.3,hatchsep=1.5,shadowsize=1,dimen=middle}
\psset{dotsize=0.7 2.5,dotscale=1 1,fillcolor=black}
\psset{arrowsize=1 2,arrowlength=1,arrowinset=0.25,tbarsize=0.7 5,bracketlength=0.15,rbracketlength=0.15}
\begin{pspicture}(0,0)(54.08,27.96)
\psline(18.16,18.38)(8.58,26.76)
\psline[linewidth=0.2](30.13,27.96)(22.95,8.8)
\psline[linewidth=0.2](30.13,27.96)(37.32,8.8)
\psline(42.11,18.38)(51.68,26.76)
\psline[linewidth=0.2](42.11,18.38)(30.13,27.96)
\psline[linewidth=0.2](18.16,18.38)(30.13,27.96)
\psline[border=0.3](37.32,8.8)(54.08,17.18)
\psline[border=0.3](22.95,8.8)(6.18,15.99)
\rput{0}(25.16,15){\psellipse[fillstyle=solid](0,0)(1.05,1.05)}
\rput{90}(38,21.9){\psellipse[fillstyle=solid](0,0)(1.05,1.05)}
\rput{0}(22.79,22.04){\psellipse[fillstyle=solid](0,0)(1.05,1.05)}
\rput{0}(34.98,15.06){\psellipse[fillstyle=solid](0,0)(1.05,1.05)}
\rput{0}(47.02,22.74){\psellipse[fillcolor=lightgray,fillstyle=solid](0,0)(1.05,1.05)}
\rput{0}(47.24,13.71){\psellipse[fillcolor=lightgray,fillstyle=solid](0,0)(1.05,1.05)}
\rput{90}(12.91,22.98){\psellipse[fillcolor=lightgray,fillstyle=solid](0,0)(1.06,1.05)}
\rput{0}(12.86,13.09){\psellipse[fillcolor=lightgray,fillstyle=solid](0,0)(1.06,1.05)}
\rput(38.97,25.03){\tiny$y_{5}$}
\rput(21,25){\tiny$y_{2}$}
\rput(27,12){\tiny$y_{3}$}
\rput(34,12){\tiny$y_{4}$}
\rput(50,12){\tiny$z_{3}$}
\rput(50,21){\tiny$z_{4}$}
\rput(10,12){\tiny$z_{2}$}
\rput(10,21){\tiny$z_{1}$}
\end{pspicture}
\end{center}
\end{figure}

\vspace{-1cm}

We now address the concrete question of finding the projector $P_{BC}$  such that
$\langle \tau_4,P_{BC}\tau_1 \rangle$ is, up to some eventual irrelevant normalization factors,
the vertex amplitude $V_{BF}$. Of course, we have implicitely assumed that $\tau_1$ and $\tau_4$ are spin-network states
i.e. there are associated to a colorating of the edges and the vertices of their associated graphs. 
Concerning $\tau_1$, its vertex is colored with an intertwiner denoted $\omega_1$ and each edge associated to
the variables $y_k$ are colored with a representation denoted $J_k$. Concerning $\tau_4$, its vertices $i$ are colored
with intertwiners $\omega_i$ and each edge associated to the variables $x_{ij}$ are colored with representations $I_{ij}$.

The operator $P_{BC}$ has to be a discretization of the flatness condition on the connection: it 
is a cylindrical distribution on $\Gamma$ which 
imposes that the holonomies around the closed faces of $\Gamma$ are trivial. One candidate which realizes such a 
requirement is given by:
\be
P_{BC} \; = \; \delta(x_{123}) \delta(x_{234}) \delta(x_{345})  \delta(x_{451})
\delta(x_{512}) 
\ee
with the notation of (\ref{definition of C}). We need only five delta distributions to impose
the flatness condition on the ten faces of the 4-simplex. Furthermore, we see that $P_{BC}$ is nothing
but the distribution $C(x_{ij})$  we have previously introduced (\ref{definition of C}).
To show that this operator is indeed a solution of our problem, let us compute
its matrix element between the states $\tau_1$ and $\tau_4$ making use of the kinematical scalar product:
\ba
\la \tau_4,P_{BF}\tau_1 \ra & = & \int (\prod_{i\neq j}^5 dx_{ij}) \, (\prod_{k=2}^5 dy_k) \, (\prod_{\ell=1}^4 dz_\ell)
\, \overline{\tau_4(x,z_\ell)} \, C(x,y) \tau_1(y,z) \nonumber \\
& = & \prod_{k=1}^4 \frac{\delta_{J_k, I_{1k}}}{d_{I_{1k}}} \, \int (\prod_{i\neq j} dx_{ij}) \, C(x_{ij}) \, \tau_5(x_{ij})
\label{PBF}
\ea
where $\tau_5(x_{ij})$ is the spin-network state associated to the 4-simplex graph. To obtain this result,
we have performed the integration over the $z$ variables first, then we have absorbed the $y$ variables  using
the invariance of the Haar measure to get as a final result an integral involving only the variables $x_{ij}$. 
At this point, it is immediate to see that the previous integral simplifies and we have:
\be
\la \tau_4,P_{BF}\tau_1 \ra \; = \; \left(\prod_{k=1}^4 \frac{\delta_{J_k, I_{1k}}}{d_{I_{1k}}}\right)
 \, V_{BF}(I_{ij},\omega_i) \;.
\ee 
Up to a renormalization factor, the physical scalar product gives exactly the desired vertex amplitude of the 
topological model. Therefore, we found a projector $P$ into the physical states of the topological model.

Let us finish the study of this case with some remarks. First, the construction of $P_{BC}$ can be easily generalized
to the space of all cylindrical functions: we only have
to impose the flatness condition aroung the closed loops of the spin-netwoks but taking into account the fact that
one has to avoid redundant delta distributions in order to have a finite amplitude. 
Second, as we have already said, the projector $P_{BC}$ has a clear physical interpretation in the sense
that it is a discretization of the first class constraints of the BF theory. For that reason, one can suppose
that the solution we found is unique. As a final remark, let us emphasize that,
even if the topological model is not physically interesting, it will appear very useful to understand the 
gravitational models,
namely the BC and the EPR models. Indeed, the three models admit the same kinematical Hilbert space
and, as we will see, the operators $P_{BC}$ and $P_{EPR}$ are constructed from the 
operator $P_{BF}$ we have just constructed. In other words, the physical scalar products of the gravitational
models are obtained from the physical scalar product of the topological model. This aspect will be precisely
described in next Section.

\subsection{The Barrett-Crane model}
This Section is devoted to the construction of the operator $P_{BC}$. For that purpose, we use the same notations
as in the previous Section concerning the space of cylindrical functions $\text{Cyl}(\widetilde{\Gamma})$,
in particular concerning the states $\tau_1$ and $\tau_4$. This makes sense because the topological and the BC
models possess the same kinematical Hilbert space.
Thus, we look for an operator $P_{BC}$ acting on the space of cylindrical functions
$\text{Cyl}(\widetilde{\Gamma})$ such that
\ba\label{defdePBC}
\la \tau_4,P_{BC} \tau_1 \ra \; = \; N \, \left(\prod_{k=1}^4 \delta_{J_k,I_{1k}}\right)  V_{BC}(I_{ij},\omega_{BC})
\ea
where $N$ is an eventual normalization factor.
We propose a solution where the projector is the product $P_{BC}=P_{BF} \tilde{P}_{BC}$
of the projector $P_{BF}=C(x_{ij})$ of the topological model and  another operator
$\tilde{P}_{BC}$ we are going to define. First, $\tilde{P}_{BC}$ has a non-trivial action on $\text{Cyl}(\Gamma)$
but can be trivially extended to the space $\text{Cyl}(\widetilde{\Gamma})$. Then,
its action on any function $F \in \text{Cyl}(\Gamma)$ is explicitely given by:
\be
(\tilde{P}_{BC} F)(x_{ij}) \; = \; \int (\prod_{i<j} dv_{ij}) \, F(v_{ij}x_{ij}v_{ij}^{-1}) 
\ee
where we used the obvious notation $x_{ij}$ for the group variable associated to the oriented edge $(ij)$
of $\Gamma$. Thus, $\tilde{P}_{BC}$ acts non-trivially on the internal edges of $\tilde{\Gamma}$; this action can 
be graphically represented as follows:
\be
\ifx\JPicScale\undefined\def\JPicScale{1}\fi
\psset{unit=\JPicScale mm}
\psset{linewidth=0.3,dotsep=1,hatchwidth=0.3,hatchsep=1.5,shadowsize=1,dimen=middle}
\psset{dotsize=0.7 2.5,dotscale=1 1,fillcolor=black}
\psset{arrowsize=1 2,arrowlength=1,arrowinset=0.25,tbarsize=0.7 5,bracketlength=0.15,rbracketlength=0.15}
\begin{pspicture}(0,0)(94,15)
\psline(10,10)(30,10)
\rput{0}(20,10){\psellipse[fillstyle=solid](0,0)(2,2)}
\rput(20,5){$x_{ij}$}
\rput(4,10){$\tilde{P}$}
\rput(34,10){$=$}
\psline(48,10)(68,10)
\rput{0}(58,10){\psellipse[fillstyle=solid](0,0)(2,2)}
\rput(58,5){$x_{ij}$}
\rput(41,10){$\int\;d v_{ij}$}
\rput{0}(65,10){\psellipse[fillcolor=gray,fillstyle=solid](0,0)(1,1)}
\rput{0}(51,10){\psellipse[fillcolor=gray,fillstyle=solid](0,0)(1,1)}
\rput(65,15){$v_{ij}^{-1}$}
\rput(51,15){$v_{ij}$}
\rput(71,10){$=$}
\rput(100,10){$d_{I_{ij}}^{-1}$.}
\psline(74,15)(94,15)
\rput{0}(84,10){\psellipse[fillstyle=solid](0,0)(2,2)}
\psecurve(82,10)(82,10)(82,10)(76,12)(84,14)(92,12)(86,10)(86,10)(86,10)
\rput(84,5){$x_{ij}$}
\end{pspicture}
\ee
Let us now see that $P_{BC}$ reproduces the physical scalar product in the sense of the equation
(\ref{defdePBC}).  Indeed, an immediate calculation leads to the result:
\ba
\la \tau_4,P_{BC} \tau_1\ra & = & \left(\prod_{k=1}^4 \frac{\delta_{J_k,I_{1k}}}{d_{I_{1k}}} \right)
\int (\prod_{i<j} dx_{ij})\, (P_{BC} \tau_5)(x_{ij}) \\
& = &  \left(\prod_{k=1}^4 \frac{\delta_{J_k,I_{1k}}}{d_{I_{1k}}} \right) 
\tau_5(1) \, \int (\prod_{i\neq j} dx_{ij})\, \frac{\chi_{I_{ij}}(x_{ij})}{d_{I_{ij}}} \, C(x_{ij})
\ea
where $\tau_5(1)$ is the spin-network $\tau_5$ evaluated at the identity $x_{ij}=1$, then it is 
the vertex amplitude of the topological model, i.e. a $SU(2)$ 15j symbol. Thus, the
previous equation can recasted the as follows:
\be\label{BCvsBF}
\frac{\la \tau_4,P_{BC} \tau_1\ra}{\la \tau_4,P_{BF} \tau_1\ra} \; = \; 
\left(\prod_{k=1}^4 \frac{\delta_{J_k,I_{1k}}}{d_{I_{1k}}} \right) 
V_{BC}(I_{ij},\omega_i^{BC}) \;.
\ee
Up to some normalization factor, the operator $P_{BC}$ reproduces the vertex amplitude of the BC model.
Thus, $P_{BC}$ can be interpreted as a projection into physical states of the BC model.

The construction we are proposing rises many important remarks.

\medskip

{\bf Remark 1.}  The operators $P_{BF}$ and $\tilde{P}_{BC}$ do not commute and therefore the order of their 
product clearly matters. The operator $P_{BF}$ is a multiplicative operator that impose the discrete analoguous
of the flatness of the Ashtekar connection and then it can be interpreted as a projector into space-diffeomorphism
invariant states. This interpretation is based on the fact that, in three dimensions, the flatness constraint on
the connection generates diffeomorphisms.
The operator $\tilde{P}_{BC}$ is a kind of ``derivative" operator
for its action involves $SU(2)$ right and left derivatives. Its physical interpretation is not clear.

\medskip

{\bf Remark 2.} If one believes that the BC model is related to gravity, then it is clear that ${P}_{BF}$
is the projection into ${\cal H}_{diff}$ and $\tilde{P}_{BC}$ should contain the projection into the kernel of
the Hamiltonian constraint.
This is far from being obvious and that conjecture is even false if the BC model is not the one that discretizes gravity
as it is suspected. Let us notice that, in our construction, $\tilde{P}_{BC}$ acts first and then acts $P_{BC}$ which is
contrary to what one usually does in LQG where the projection into ${\cal H}_{diff}$ arises before the projection
into the kernel of the Hamiltonian constraint.

\medskip

{\bf Remark 3.} The operator $\tilde{P}_{BC}:\text{Cyl}(\Gamma) \rightarrow C(SU(2))_{Ad}^{\times 10}$ is in fact
a projector from the space of cylindrical functions to ten copies of the space of functions on the conjugacy classes 
$C(SU(2))_{Ad}$ of the group $SU(2)$ where $F \in C(SU(2))_{Ad}$ if and only if $F(gxg^{-1})=F(x)$ for any $x$ and
$g$ in $SU(2)$. Its action on a $\tau_5$ spin-network state is given by:
\be
(\tilde{P}_{BC} \tau_5)(x_{ij}) \; = \; \tau_5(1) \, \prod_{i<j} \frac{\chi_{I_{ij}}(x_{ij})}{d_{I_{ij}}} 
\ee
where $I_{ij}$ are the representations coloring the edges $(ij)$ of the graph $\Gamma$.
It is straightforward to check that $\tilde{P}_{BC}^2=\tilde{P}_{BC}$. 
As a consequence, for the definition of ${P}_{BC}$ to make sense, one has to extend $P_{BF}$ as an operator acting on 
$C(SU(2))_{Ad}^{\times 10}$ which is trivial.

\medskip

{\bf Remark 4.} In fact, the decomposition of $P_{BC}$ as the product of $P_{BF}$ and $\tilde{P}_{BC}$ is not canonical.
Our construction provides an equivalent class of functions $\tilde{P}_{BC}$ according to the trivial relation 
$\tilde{P}_{BC} \sim \tilde{Q}_{BC}$ if and only if $P_{BF} \tilde{P}_{BC}  = P_{BF} \tilde{Q}_{BC}$.
Another natural choice for the derivative operator is $\tilde{Q}_{BC}$ defined by its
following action on $\tau_5$ spin-network states: 
\be
(\tilde{Q}_{BC} \tau_5)(x_{ij}) \; \equiv \; \tau_5(x_{ij}) \prod_{i<j} \frac{\chi_{I_{ij}}(x_{ij})}{d_{I_{ij}}} \;.
\ee
This representative is clearly a multiplicative operator.

\medskip

{\bf Remark 5.} As a last remark, let us underline that the physical scalar product between two states
in the BC model (\ref{BCvsBF}) can  be viewed as the matrix element of the operator $\tilde{P}_{BC}$ 
with respect to the physical scalar product of the topological model up to the ``norm"
$\langle \tau_4, P_{BF} \tau_1 \rangle$. In that sense, the BC model is very closely related to
the topological model.

\subsection{The Engle-Pereira-Rovelli model}
In this Section,
we propose an operator $P_{EPR}$ which reproduces the vertex amplitude of the EPR model.
The construction of $P_{EPR}$ is very similar to the construction of $P_{BC}$.
As for the BC model, $P_{EPR}$ is the product of the non-commuting operators, $P_{EPR}=P_{BF}\tilde{P}_{EPR}$,
one of them being the projetor of the topological model as well.
The operator $\tilde{P}_{EPR}$ is defined by its action on spin-network states $\tau_5(x_{ij})$ explicitely
given by:
\ba\label{intVepr}
(\tilde{P}_{EPR} \tau_5)(x_{ij}) \; = \; \int (\prod_{i <j} dv_{ij} dv_{ji}) \, 
\left[\prod_{i<j} \chi_{\tilde{I}_{ij}}(v_{ij}x_{ij}v_{ji}) \chi_{\tilde{I}_{ij}}(v_{ij}) 
\chi_{\tilde{I}_{ij}}(v_{ji})\right] \, \tau_5(v_{ij}x_{ij}v_{ji})
\ea
where we have introduced the notation $\tilde{I}_{ij}=I_{ij}/2$. As in the BC model, $\tilde{P}_{EPR}$
acts on each edge of the spin-network and this action can be pictured as follows:
\be
\ifx\JPicScale\undefined\def\JPicScale{1}\fi
\psset{unit=\JPicScale mm}
\psset{linewidth=0.3,dotsep=1,hatchwidth=0.3,hatchsep=1.5,shadowsize=1,dimen=middle}
\psset{dotsize=0.7 2.5,dotscale=1 1,fillcolor=black}
\psset{arrowsize=1 2,arrowlength=1,arrowinset=0.25,tbarsize=0.7 5,bracketlength=0.15,rbracketlength=0.15}
\begin{pspicture}(0,0)(103,32)
\psline(8,15)(28,15)
\rput{0}(18,15){\psellipse[fillstyle=solid](0,0)(2,2)}
\rput(18,10){$x_{ij}$}
\rput(2,15){$\tilde{P}_{EPR}$}
\rput(31,15){$=$}
\psline(49,15)(69,15)
\rput{0}(59,15){\psellipse[fillstyle=solid](0,0)(2,2)}
\rput(59,20){$x_{ij}$}
\rput(40,15){$\int\;d v_{ij} dv_{ji}$}
\rput{0}(66,15){\psellipse[fillcolor=gray,fillstyle=solid](0,0)(1,1)}
\rput{0}(52,15){\psellipse[fillcolor=gray,fillstyle=solid](0,0)(1,1)}
\rput(66,19){\tiny $v_{ji}$}
\rput(52,19){\tiny $v_{ij}$}
\rput(73,15){$=$}
\rput{0}(90,13){\psellipse[fillstyle=solid](0,0)(2,2)}
\psecurve(88,13)(88,13)(88,13)(82,15)(90,17)(98,15)(92,13)(92,13)(92,13)
\rput(90,9){$x_{ij}$}
\rput{0}(59,25){\psellipse[fillstyle=solid](0,0)(2,2)}
\psecurve(57,25)(57,25)(57,25)(50,27)(59,29)(68,27)(61,25)(61,25)(61,25)
\rput{0}(52,25){\psellipse[fillcolor=gray,fillstyle=solid](0,0)(1,1)}
\rput{0}(66,25){\psellipse[fillcolor=gray,fillstyle=solid](0,0)(1,1)}
\rput{0}(52,4.5){\psellipse[](0,0)(3,-1.5)}
\rput{0}(52,6){\psellipse[fillcolor=gray,fillstyle=solid](0,0)(1,1)}
\rput{0}(66,4.5){\psellipse[](0,0)(3,-1.5)}
\rput{0}(66,6){\psellipse[fillcolor=gray,fillstyle=solid](0,0)(1,1)}
\rput(52,10){\tiny $v_{ij}$}
\rput(66,10){\tiny $v_{ij}$}
\psline(78,15)(82,15)
\psline(98,15)(102,15)
\rput(11,17){$I_{ij}$}
\rput(25,17){$I_{ij}$}
\rput(59,32){$\frac{I_{ij}}{2}$}
\rput(66,0){$\frac{I_{ij}}{2}$}
\rput(52,0){$\frac{I_{ij}}{2}$}
\rput(103,17){$I_{ij}$}
\rput(90,20){$\frac{I_{ij}}{2}$}
\rput(77,17){$I_{ij}$}
\end{pspicture}
\ee
In this figure, the closed loops represent $SU(2)$ characters. The last equality has been obtained after integrating
over the $v_{ij}$ variables.
Using this pictorial representation, it is quite easy to compute the matrix elements of $P_{EPR}$
between the states $\tau_1$ and $\tau_4$:
\ba
\la \tau_4,P_{EPR} \tau_1\ra & = & \left(\prod_{k=1}^4 \frac{\delta_{J_k,I_{1k}}}{d_{I_{1k}}} \right)
\int \prod_{i<j} dx_{ij} \, (P_{EPR} \tau_5)(x_{ij}) \\
& = & \left(\prod_{k=1}^4 \frac{\delta_{J_k,I_{1k}}}{d_{I_{1k}}} \right) \,N^{-1} \,
V_{EPR}({I}_{ij},\omega_i^{EPR}) 
\ea
where $N$ is the normalization factor introduced in (\ref{simplifiedintegralEPR}).
As a consequence, we claim that the physical scalar product reproduces the vertex amplitude of the EPR
model. Let us now finish this Section with some important remarks.

\medskip

{\bf Remark 1.} The previous remarks 1 and 2 concerning the BC model can be transposed to the EPR model.
In particular, if the EPR model is a discretization of the path integral of gravity, $\tilde{P}_{EPR}$
should be closely related to the Hamiltonian constraint. If this is true, our formula could give some hints
about the regularization of the Hamiltonian constraint. Furthermore, we can easily generalize the construction
to any cylindrical functions with no restriction on the underlying graph.

\medskip

{\bf Remark 2.} The operator $\tilde{P}_{EPR}$ is constructed making use of an integration over 20
variables $v_{ij}$ with $i\neq j$ because $v_{ij}\neq v_{ji}$ in the formula (\ref{intVepr}). Contrary to the BC
model, $\tilde{P}_{EPR}$ is not a projector neither an operator from $\text{Cyl}(\Gamma)$ to the space of 
functions on the $SU(2)$ conjugacy classes. The integral (\ref{intVepr}) can be reduced to an integral
over only 10 variables $v_{ij}$ with $i<j$ as follows:
\be
(\tilde{P}_{EPR} \tau_5)(x_{ij}) \; = \; \int (\prod_{i<j} dv_{ij}) 
\left[ \prod_{i<j} \frac{\chi_{\tilde{I}_{ij}}(v_{ij}) \chi_{\tilde{I}_{ij}}(x_{ij}^{-1}v_{ij})}{d_{\tilde{I}_{ij}}}\right] \tau_5(v_{ij}) \;.
\ee
To obtain such a formula, we have first integrated over the variables $v_{ij}$ with $i>j$ and then we have performed
some changing of variables. 

\medskip

{\bf Remark 3.} Our construction can be generalized immediately to the FK models presented in Section 1.3.4. 
The resulting operator $P_{FK}$ would take exactly the same form as $P_{EPR}$ with some differences in the representations
of the characters in the integrand of (\ref{intVepr}).

\medskip

{\bf Remark 4.} Concerning the unicity of $\tilde{P}_{EPR}$, we can make the same remark 4 as in the BC model,
namely our construction provides a certain equivalent class of solutions for $\tilde{P}_{EPR}$ and the decomposition
of $P_{EPR}$ as a product of $P_{BF}$ and $\tilde{P}_{EPR}$ is not canonical.

\section*{Conclusions and perspectives}
On the first hand, 
this article opens one way towards the understanding of an eventual link between Loop Quantum Gravity and
Spin-Foam models.  We have shown that the vertex amplitudes of some Spin-Foam models can be precisely interpreted
as a ``physical" scalar product between two spin-networks. This makes a clear relation between the canonical
and covariant quantizations of 4 dimensional Euclidean gravity. It is indeed possible to construct operators
$P$ acting on the space 
$\text{Cyl}(\Gamma)$ 
of cylindrical functions on the (extended) 4-simplex graph $\tilde{\Gamma}$
such that its matrix elements between spin-networks states gives, up to some eventual normalization, the vertex
amplitudes for Spin-Foam models. In a formal language, we have shown that
\be\label{conclusion}
\la s , P s' \ra \; = \; {\cal A}(s,s')
\ee
where $\la,\ra$ is the kinematical scalar product; $s$ and $s'$ belongs to ${\text{Cyl}}(\tilde{\Gamma})$
and ${\cal A}(s,s')$ is the Spin-Foam amplitude of a graph interpolating between $s$ and $s'$ which is, here,
proportional to the vertex amplitude.
The construction works for the topological model, the Barrett-Crane model,
the Engle-Pereira-Rovelli model and their direct generalizations, namely the Freidel-Krasnov models.

\medskip 

On the other hand, the same article opens questions that certainly deserve to be investigated.
The first one concerns the possibility to extend our construction to the case where the spin-networks
$s$ and $s'$ (\ref{conclusion}) are any cylindrical functions and not restricted to $\text{Cyl}(\tilde{\Gamma})$ as this 
was the case in this article. It is clear that the action of the operators $P$ we have constructed can be 
easily extended to any spin-networks with no assumption on the underlying graph defining the spin-networks.
It would be very nice to first compute the matrix elements of $P$ between these general states and to check if
the result is related to a Spin-Foam amplitude associated to a graph interpolating between the two
associated spin-network graphs. We hope to study this very exciting problem in the close future.

The second question concerns the link between the operators $P$ we have constructed and the 
regularization of the Hamiltonian constraint \`a la Thiemann. Indeed, one would expect that,
if the Spin-Foam models are a discretized version of the path integral of gravity, then $P$
should be related to the Hamiltonian constraint. It is interesting to remark that the projector $P_{EPR}$
of the $EPR$ model have some ``similarities" with the Thiemann constraint: for example, it acts on the
nodes of the spin-networks. It is nonetheless intriguing to notice an important difference between the ways 
the constraints are imposed in LQG and in the Spin-Foam models throught the operators $P$: 
indeed, in LQG, one imposes the vectorial
constraint before imposing the scalar constraint whereas the operator $P=P_{BF}\tilde{P}$ 
is the non-commutative product of two
operators, the second one $P_{BC}$ imposes clearly the space-diffeomorphisms invariance and ``projects" 
into the vectorial constraint kernel. Of course, it is too early to conclude anything but its seems to have
a quite important discrepencie between the two approaches. To understand more precisely these aspects, one
could start by understanding the link between the projector $P$ and the classical constraints of gravity.

The third question is more mathematical: is $P$ a GNS state? Indeed, it is quite misleading to view
$P$ as an operator acting on cylindrical functions for it is a linear form on $\text{Cyl}(\tilde{\Gamma})$.
Thus it seems that the GNS theory is the good mathematical framework to study $P$. But, if one wants to
interpret $P$ has a GNS state, one has to check that it satisfies all the required property, among other 
the positivity. 

We finish this conclusion by mentionning the possibility that our work could give some hints
to study the classical and semi-classical behaviors of the EPR model.

\subsubsection*{Aknowledgments}
We are very grateful to Carlo Rovelli for many things: his interest and his enthusiasm to this subject,  
his reading of the paper and for leading us to study this problem.
The work was partially supported by the ANR (BLAN06-3\_139436 LQG-2006).
E.Alesci wish to thank the support by
Della Riccia Foundation.

\end{document}